\documentclass[draftcls,12pt,onecolumn]{IEEEtran}
\usepackage{setspace}
\usepackage{amsmath,graphicx}
\allowdisplaybreaks[4]
\usepackage{amssymb}
\usepackage{xcolor}
\usepackage{array}
\usepackage{relsize}
\usepackage{eqparbox}
\usepackage{mdwmath}
\usepackage{mdwtab}
\usepackage{mathtools}
\usepackage{wrapfig}
\usepackage[justification=centering]{caption}

\newcommand{\beq}{\begin{equation}}
\newcommand{\eeq}{\end{equation}}
\newcommand{\beqa}{\begin{eqnarray}}
\newcommand{\eeqa}{\end{eqnarray}}
\newcommand{\beqs}{\begin{equation*}}
\newcommand{\eeqs}{\end{equation*}}
\newcommand{\beqas}{\begin{eqnarray*}}
\newcommand{\eeqas}{\end{eqnarray*}}

\newcommand{\nnb}{\nonumber}
\renewcommand\thesubsection{\thesection.\alph{subsection}}
\def\conv{\mathbin{\hbox{$\,*$\kern-1.5ex$\circledast$}}}
\newcommand{\rref}[1]{Fig. \ref{#1}}

\title{An efficient time domain representation for Single-Carrier Frequency Division Multiple Access}
\author{\IEEEauthorblockN{Bouchra Benammar$^*$, Nathalie Thomas$^*$,  Marie-Laure Boucheret$^*$, Charly Poulliat$^*$, and Mathieu Dervin$^{\dag}$\\}
\IEEEauthorblockA{$^*$ University of Toulouse, INPT-ENSEEIHT/IRIT\\ $^\dag$ Thales Alenia Space, Toulouse\\ 
Email: {\{bouchra.benammar, nathalie.thomas, marie-laure.boucheret, charly.poulliat\}@enseeiht.fr, \\ mathieu.dervin@thalesaleniaspace.com}}}

\begin{document}
\maketitle
\begin{abstract}
This paper presents a physical model for Single Carrier-Frequency Division Mutliple Access (SC-FDMA). We specifically show that by using mutlirate signal processing we derive a general time domain description of Localised SC-FDMA systems relying on circular convolution.
This general model has the advantage of encompassing different implementations with flexible rates as well as additional frequency precoding such as spectral shaping. Based on this time-domain model, we study the Power Spectral Density (PSD) and the Signal to Interference and Noise Ratio (SINR). Different implementations of SC-FDMA are investigated and analytical expressions of both PSD and SINR compared to simulations results.
\end{abstract}

\section{Introduction}\label{sec:intro}
 
The Third Generation Partnership Project (3GPP) Long Term Evolution (LTE) is a radio standard adopted for new generation mobile networks in order to put up with the increasing demand for high data rates. Increased data rates lead to an increased frequency selectivity of the channel which could be mitigated by using multi-carrier transmissions. Thus, the LTE proposed physical layer uses Orthogonal Frequency Division Multiple Access (OFDMA) in the downlink. However, the standard elected SC-FDMA as the uplink technology, since among other advantages, it experiences less Peak to Average Power Ratio (PAPR). Indeed, SC-FDMA initially proposed in \cite{Myung_Mag_06} consists of a DFT precoded OFDMA, which explains its low PAPR compared to OFDMA. The PAPR reduction is of paramount importance when it comes to the energy efficiency of the Power Amplifiers (PA) embedded in User Equipments (UE). In fact, in order not to distort the amplified signal, PAs need to be operated with a back-off toward their input saturation power. This back-off is all the more significant when the input signal has large dynamics i.e. high PAPR. 
Increasing the amplifiers back-off, lowers their power efficiency and shortens the UEs' battery life. Thanks to the DFT-precoding, SC-FDMA (also referred to as DFT-spread OFDMA) has single carrier characteristics in terms of signal dynamics and thus has lower PAPR than OFDMA which explains why it has been proposed as the uplink transmission scheme. \\
Two different mapping schemes of SC-FDMA have been suggested, differing in the way users are multiplexed into the available sub-carriers. A first mapping denominated Localised FDMA (LFDMA) consists of allocating contiguous blocks of sub-carriers to each user. The second category assigns a chunk of evenly spaced sub-carriers to each user and is thus called Distributed FDMA (DFDMA). A special case of DFDMA allocates only one subcarrier at a regular spacing for each user and is called Interleaved FDMA (IFDMA). Original study of IFDMA has been presented in \cite{Frank} and shows that it is equivalent to compressing and repeating the input users symbols in the time domain. As such, IFDMA has lower PAPR than LFDMA. However, since a fine users synchronisation is required for IFDMA, Carrier Frequency Offsets (CFO) \cite{Zhang2007} and phase noise \cite{Sridharan_2012} can have significant impact on the system performance. This explains why LFDMA has been preferred to IFDMA in the LTE standard. \\
A further reduction of the PAPR still remains desirable for SC-FDMA. Many of the OFDMA PAPR reduction techniques can be applied to SC-FDMA as a special case of precoded OFDMA \cite{Seung_2005} \cite{Lemoine_2007}. Among the family of time domain solutions, authors in \cite{Meza_2008} proposed parametric linear pulse shapes which are Nyquist-shapes having lower PAPR than raised cosine pulse shaping. \\
The second family relies on frequency domain precoding or spectral shaping \cite{Slimane_2007}, \cite{Falconer_2011}, and \cite{Behrouz_2012}. In \cite{Slimane_2007}, raised cosine frequency shapes were investigated at the cost of decreased spectral efficiency. In \cite{Falconer_2011}, Falconer presented a linear frequency precoding window which (numerically) minimizes the variance of the instantaneous output power but induces a slight noise enhancement. In contrast, authors in \cite{Behrouz_2012} proposed a mathematical model of the PAPR reduction problem and derived new optimized windows using Langrangian multipliers. These frequency windows optimized with the Compensation of Noise Enhancement Penalty (CNEP) reduce the PAPR and improve the system performance in terms of Bit Error Rate (BER).  \\ 
What can be noticed throughout these works is the lack of a unified model of the SC-FDMA system, some considering it as a precoded OFDMA, some as a special case of generalised multicarrier system \cite{Lemoine_2007}. In \cite{Viholainen_2009}, authors proposed an efficient scheme to generate SC-FDMA comparable waveforms called SciFI-FDMA. Frequency and time domain interpolations followed by frequency shifting are used to reduce the complexity of a classical SC-FDMA scheme due to the flexible size $K$ of the precoding DFT. The scheme has been proved less complex but suffers from approximation errors due to interpolation. In this article, we show that for study purposes, SC-FDMA can be efficiently modelled as a single carrier transmission scheme by giving a time domain model relying on circular convolution which encompasses different linear frequency precoding schemes. Since the equivalent time domain representation of Interleaved FDMA has already been developed in \cite{Frank}, this paper will only be interested in Localised SC-FDMA.  \\ 
Besides, we will also be interested in some system design aspects and more specifically the Power Spectral Density (PSD) and the Signal to Interference Noise Ratio (SINR). 
PSD analysis is an essential feature to ensure that the transmit power spectrum is confined within a predefined transmission spectrum mask. It is also valuable for resource allocation among different users \cite{babar_2011}. 
Unlike PAPR reduction techniques, OFDMA PSD formulas  \cite{Waterschoot_2010} can not be directly applied to SC-FDMA since DFT precoding changes the statistical properties of the OFDMA input symbols. In a previous work \cite{eusipco_bouchra}, we proposed analytical expressions of the SC-FDMA PSD with general spectral shaping relying on the frequency domain representation of SC-FDMA. In this work, we  derive PSD formulas based on the novel time domain representation of SC-FDMA and apply it to different versions of spectrally shaped SC-FDMA implementations. \\
We will show that the new time domain model leads to a convenient and simple derivation of the SINR with linear equalizers that can be applied to any LFDMA scheme. The advantage of this analytical result is that it applies to a wide range of frequency precoding schemes as well as fractional and non fractional SC-FDMA rates. Knowing the SINR is an essential feature for Bit Error Rate prediction methods based on physical layer abstraction methods and thus for both link and system level analysis. \\
The remaining of this paper is organised as follows: after a brief presentation of the frequency based SC-FDMA model and some multirate identities, we will derive the general time domain model for Localised SC-FDMA in section \ref{sec:timedomain}. This general model takes into account the fractional rate i.e. when the OFDMA IFFT size is not multiple of the precoding FFT size. In section \ref{sec:psd}, we derive PSD formulas for the previously derived model with general spectral shapes and compare them with simulations in \ref{sec:psd_appli}. In section \ref{sec:sinr}, we derive SINR formulas for the case of Localised FDMA and compare them with empirical results from simulations on a frequency selective fading channel in section \ref{sec:lte}. Conclusions and discussions are given in section \ref{sec:concl}. \\ 
\textit{\textbf{Notations}}: In the following, the term  $A$-FFT (resp. $B$-IFFT) designates a FFT over $A$ points (resp. IFFT over $B$ points). 
Time domain (resp. frequency domain) variables are represented by lower (resp. upper) case letters.  
The FFT (resp. IFFT) operator applied to time domain symbols $x_{n}$ (resp. frequency domain symbols $X_{k}$) of length $L$ write as follows: 
\beqa
x_{n} &=& IFFT_L(X_{k})=\frac{1}{L}\sum_{p=0}^{L-1} X_{p} \Omega_L^{pn} \nnb \\
X_{k} &=& FFT_L(x_{n})=\sum_{p=0}^{L-1} x_{p} \Omega_L^{-pk} \nnb 
\eeqa 
where $\Omega_L^{pk} = e^{\frac{j 2\pi p k}{L}}$. 
The notation $(.)_{k,l}$ indicates the symbol on the $k^{th}$ sub-carrier for the $l^{th}$ time domain symbol.
\section{From frequency to time domain representation}\label{sec:timedomain}

\subsection{Frequency based SC-FDMA scheme description}\label{sec:freqdomain}

\begin{figure*}[!t]
\centering
\includegraphics[width=4in]{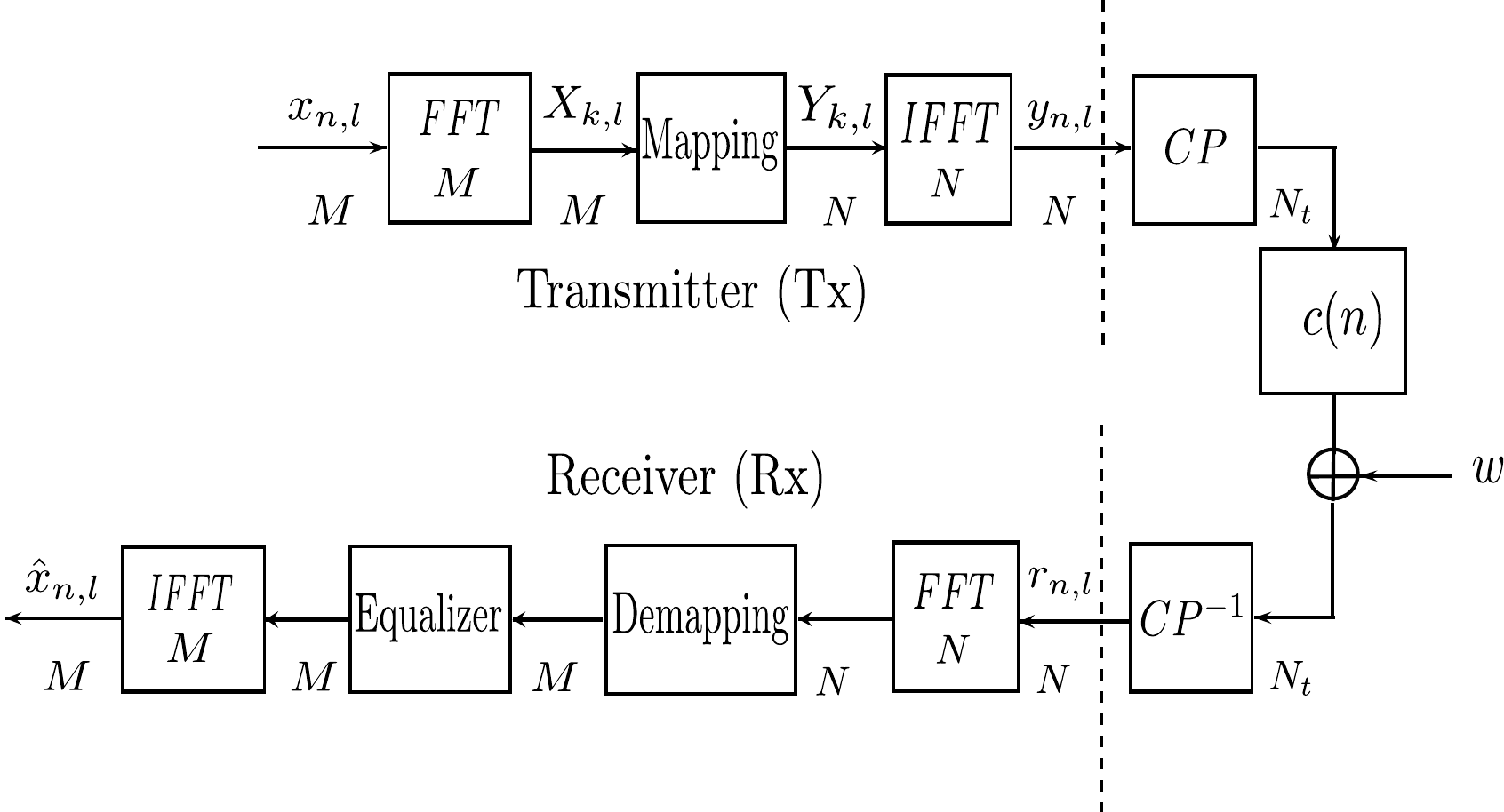}
\caption{SC-FDMA frequency based scheme representation}
\label{fig:scfdma_freq}
\end{figure*}
Consider the scheme depicted in \rref{fig:scfdma_freq}. Blocks of $M$ independent zero mean and identically distributed symbols $x_{k,n}$ at a rate $R_s$ are converted to frequency domain symbols $X_{k,n}$ by an $M$-FFT, then mapped into $M$ out of $N$ sub-carriers before being converted back to the time domain by an $N$-IFFT. In order to cope with the frequency selectivity of the channel $c(n)$, a Cyclic Prefix (CP) of length $N_g$ is appended to the resulting time domain symbols to build the SC-FDMA symbol of length $N_t = N + N_g$. This results in simplified equalization at the receiver, thanks to the circularity of the channel matrix due to the CP. Even though the CP is a part of the transmitter processing, it is integrated in the channel part in \rref{fig:scfdma_freq} since it is mostly for channel circularization. 
The signal is then affected by an additional Gaussian circular noise $w$ with variance $\sigma_w^2$. At the receiver, after CP removal, symbols are transformed into frequency domain symbols using a $N$-FFT. The sub-carriers are then demapped in order to extract the corresponding user data symbols. A frequency domain equalizer is then used to cope with channel impairments. The obtained frequency symbols are converted back to the time domain using a $M$-IFFT. 

Two principal schemes have been proposed for SC-FDMA to map the $M$ frequency symbols into the $N$ available sub-carriers, namely Localised and Interleaved mappings. 
Without loss of generality and unless otherwise stated, the user is mapped into the first block of $M$ IFFT inputs. 
In the Localised mapping, the $M$-FFT outputs $X_{k,n}$ are directly mapped into a block of contiguous $N$-IFFT inputs as follows: \\
\beqa 
Y_{k,l}^{localised} = \left \{ \begin{array}{ll}
			 X_{k,l} & if \: \: \: 0 \leq k \leq M-1\\
            0 & elsewhere 
\end{array} \right.
\eeqa
The time domain representation of SC-FDMA is interesting in the way that it would allow for a simple derivation of SNR and SINR formulas. In the originally proposed SC-FDMA, the IFFT size $N$ needs not be a multiple of $M$. The LTE fractional case i.e. $N \neq kM, \: \: k \in \mathbb{N}^*$ is intended for flexible resource allocation among users. Thus, in order to allow for general (fractional) values of $M$ and $N$, we will derive a time based SC-FDMA model using higher rates FFT/IFFT along with some repetition and overlapping operations. 
More precisely, we use FFT and IFFT operations with size equal to the Least Common Multiple (LCM) of $M$ and $N$. This allows us to derive a simple yet effective system description relying on circular convolution. 
To do so, we start by reminding some well established multi-rate noble FFT/IFFT identities.  

\subsection{\textbf{Mutli-rate FFT/IFFT noble identities}}

\begin{figure}[!t]
\centering
\includegraphics[width=3in]{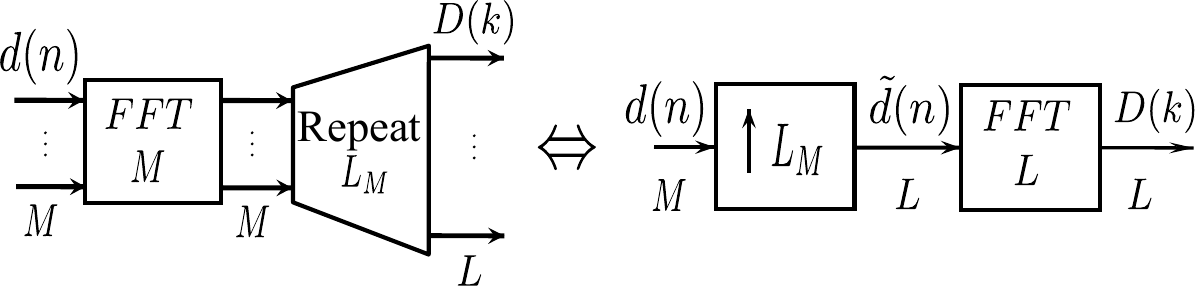}
\caption{Up-sampling identity}
\label{fig:up_sampl}
\end{figure}

\begin{figure}[!t]
\centering
\includegraphics[width=3in]{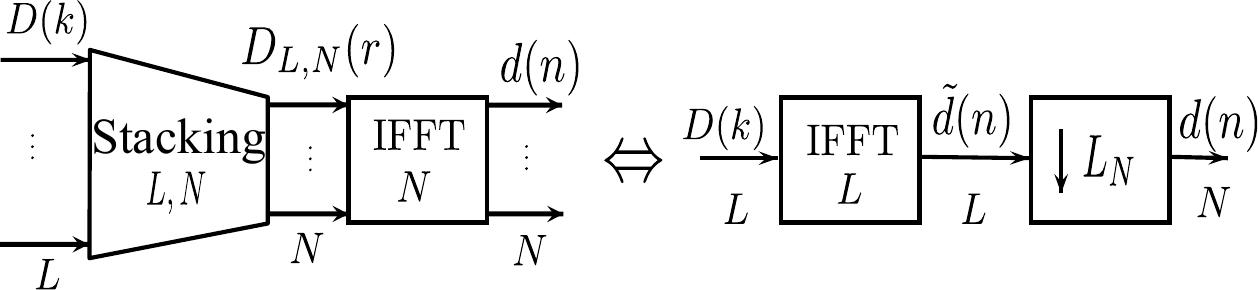}
\caption{Down-sampling identity}
\label{fig:down_sampl}
\end{figure}

\begin{figure}[!t]
\centering
\includegraphics[width=2.5in]{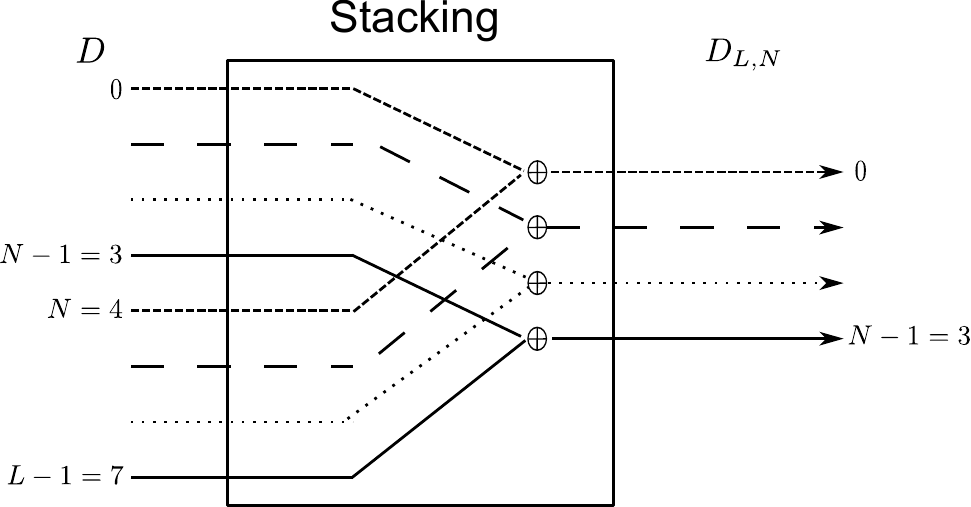}
\caption{An example of stacking with $L=8$, $N=4$, and $L_N=2$}
\label{fig:stacking}
\end{figure}

%
Let $L$ be the LCM of $M$ and $N$ i.e. $L=M L_M = N L_N $ where $L_M,\: L_N\in \mathbb{N}$. Obviously, if $N$ is a multiple of $M$ then $L=N$ and $L_N = 1$ which means that the following general results can be easily applied for the specific case when $N$ is a multiple of $M$. Moreover $L_M$ and $L_N$ are co-prime numbers i.e. they do not have any common divider and they satisfy $L_M | N$ and $L_N | M$ where "$|$" stands for "divides".\\
Let us consider the multi-rate equivalences depicted in \rref{fig:up_sampl} and \rref{fig:down_sampl} \cite{Vaidyanathan}.
\begin{itemize}

\item \textbf{Up-sampling identity}: The cascade of up-sampling by a factor $L_M$ followed by a FFT of size $L$ is equivalent to a FFT of size $M$ followed by an $L_M$-fold repetition of the $M$ outputs.
\item \textbf{Down-sampling identity}: The cascade of IFFT of size $L$ followed by $L_N$-down-sampling is equivalent to stacking with parameters $(L,N)$ followed by a IFFT of size $N$. 
Stacking with parameters $(L,N)$ consists of a summation of $L$ terms at a regular spacing equal to $N$ as depicted in \rref{fig:stacking}. This means that for $r \in {0,\hdots, N-1}$: \\
\beq
D_{L,N} (r)=  \frac{1}{L_N}  \sum_{s=0}^{L_N-1} D(sN + r)
\label{equ:stacking}
\eeq
where $L_N = \frac{L}{N}$. 
\end{itemize}

\begin{figure}[!t]
\centering
\includegraphics[width=3in]{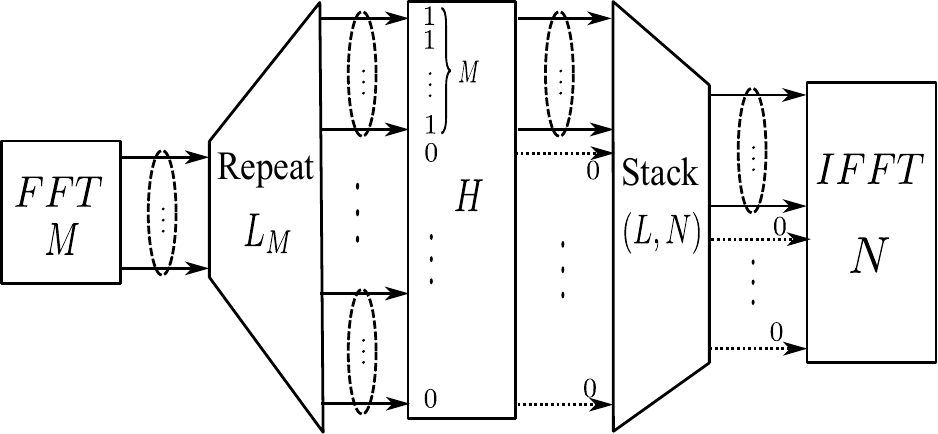}
\caption{Localised mapping modelling}
\label{fig:fig_mapping}
\end{figure}
\begin{figure}[!t]
\centering
\includegraphics[width=3.5in]{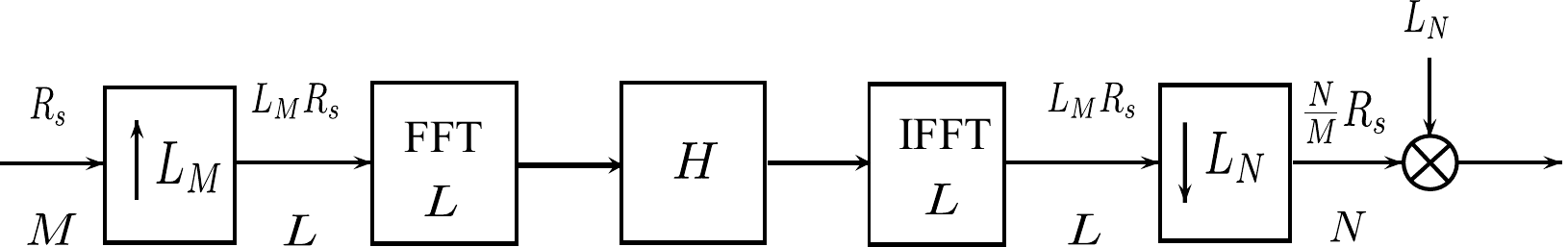}
\caption{Transmitter equivalent model}
\label{fig:Tx_model}
\end{figure}
In the following we give equivalent models for different parts of the Localised FDMA system depicted in \rref{fig:scfdma_freq} and more specifically the transmitter (Tx), the receiver (Rx), and the selective channel. 
\subsection{\textbf{Transmitter (Tx) equivalent model}} 
Let us consider the transmitter delimited by (Tx) in Fig.\ref{fig:scfdma_freq}. When using localised mapping in SC-FDMA i.e. LFDMA, the user's $M$-FFT outputs are mapped into the first block of $M$ entries in the $N$-IFFT inputs. This operation can be seen equivalent the scheme depicted in \rref{fig:fig_mapping}. The $M$ inputs are block-repeated $L_M$ times to generate $L$ inputs. The sampling rate is increased from the symbol rate $R_s$ to $L_M R_s$. The $L$ samples are then multiplied with an equivalent transmit shaping window with frequency response $H$ of length $L$ where only $M$ entries are non zero. Finally to obtain the $N$-IFFT inputs, a stacking operator following (\ref{equ:stacking}) is used in order to combine the $L$ elements into $N$ IFFT inputs. The resulting sampling frequency is equal to $\frac{N}{M} R_s$. It can be shown that for stacking with parameters $(L,N)$ if the input symbols have less than or equal to $N$ non zero inputs, then the stacking operator is only a multiplication with a factor $\frac{1}{L_N}$ of these $N$ elements. Thus, when $H$ has only $M$ non zero elements ($M \leq N$) , the stacking operator allows bringing the first user's $M$-FFT outputs to the $M$ first inputs of the $N$-IFFT, with the other $N-M$ entries equal to $0$, multiplied by $\frac{1}{L_N}$.
By using multi-rate identities, the system becomes equivalent to the model depicted in \rref{fig:Tx_model}. Multiplication with $L_N$ after the $N$-IFFT compensates for the stacking multiplication factor. 
\subsection{\textbf{Receiver (Rx) modelling}}

\begin{figure}[!t]
\centering
\includegraphics[width=3in]{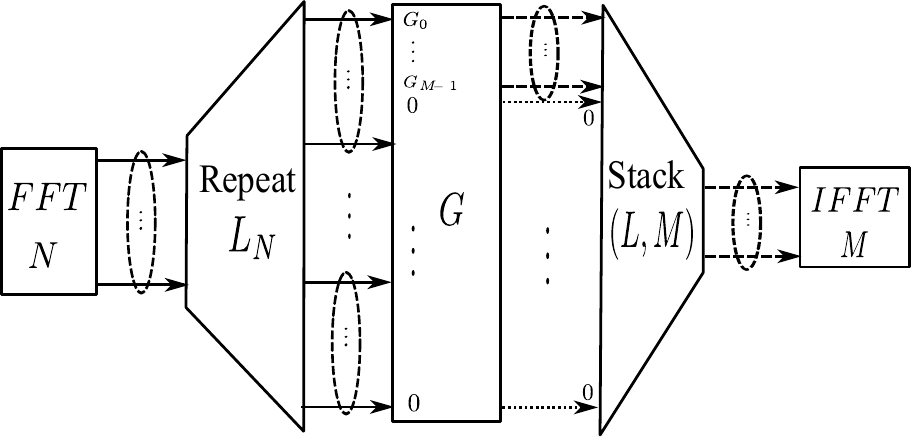}
\caption{Localised demapping and equalization equivalent model}
\label{fig:demapping_model}
\end{figure}
\begin{figure}[!t]
\centering
\includegraphics[width=3.5in]{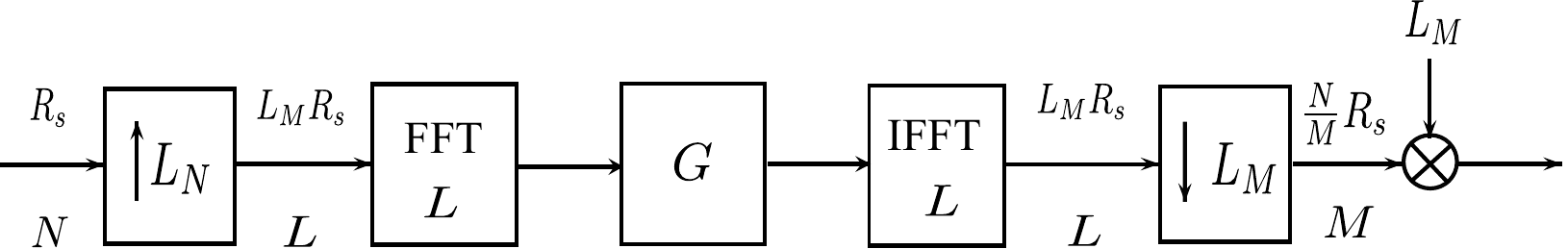}
\caption{Receiver equivalent model}
\label{fig:rx_model}
\end{figure}

At the receiver, after CP removal, the block of $N$ received samples is processed by an $N$-FFT. The user's corresponding $M$ frequency bins are extracted out of the $N$ bins through demapping. They are then equalized with a one-tap frequency domain equalizer of length $M$ thanks to the circularity of the channel. An $M$-IFFT transforms the equalized frequency samples into user's estimated time domain symbols. As with the transmitter modelling, we will use multi-rate identities to define the equivalent receiver model illustrated in \rref{fig:demapping_model}. Indeed, the receiver processing can be decomposed first into an $N$-FFT followed by a repetition of size $L_N$ leading to $L$ samples. The resulting samples are then jointly demapped and equalized using a frequency response $G$ of length $L$ which is non-null only in the user's allocated frequency bins. A stacking operator of parameters $(L,M)$ is used to combine the $L$ resulting bins into $M$ frequency symbols which are then processed by an $M$-IFFT block to yield the user's estimated symbols. By using the noble multi-rate identities, the receiver can be modelled as in \rref{fig:rx_model}. 

\subsection{\textbf{Global system time domain equivalent model}}

\begin{figure*}[!htbp]
\centering
\includegraphics[width=4.5in]{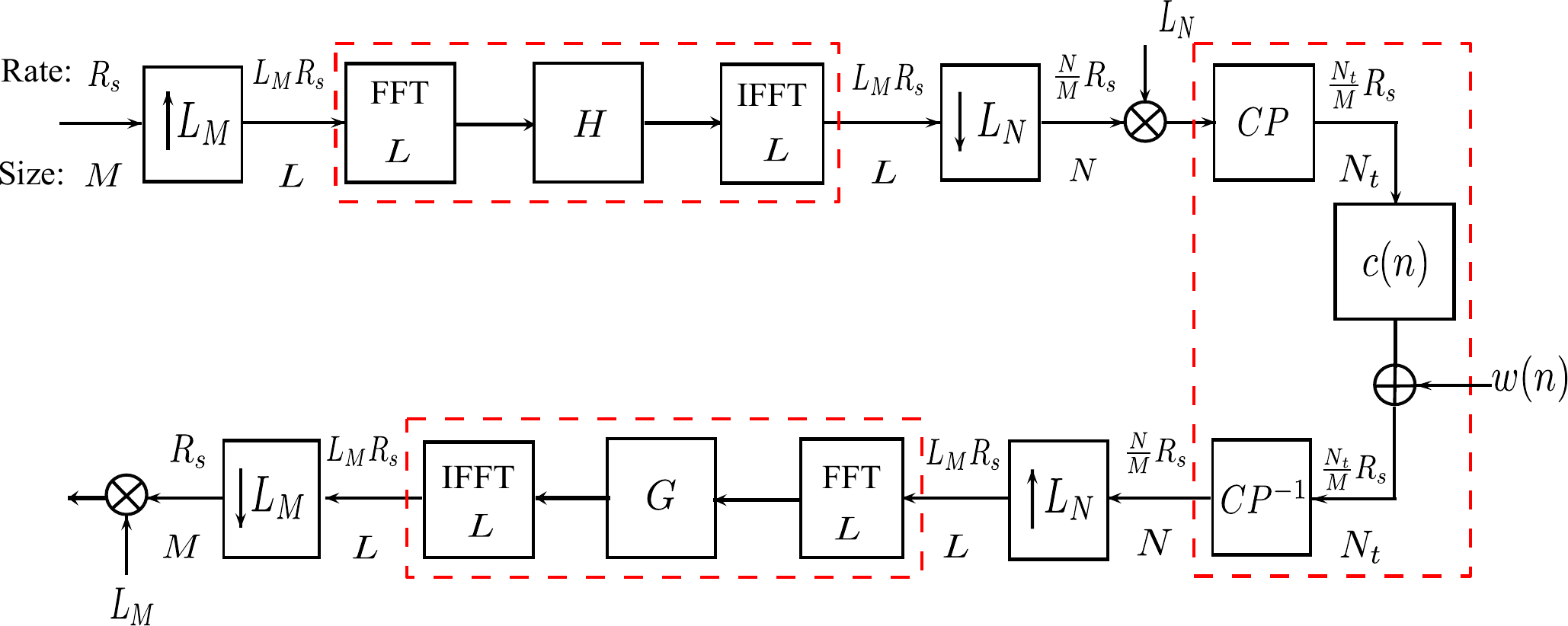}
\caption{SC-FDMA equivalent frequency domain system model}
\label{fig:figure_up_down}
\end{figure*}
\begin{figure}[!t]
\centering
\includegraphics[width=3in]{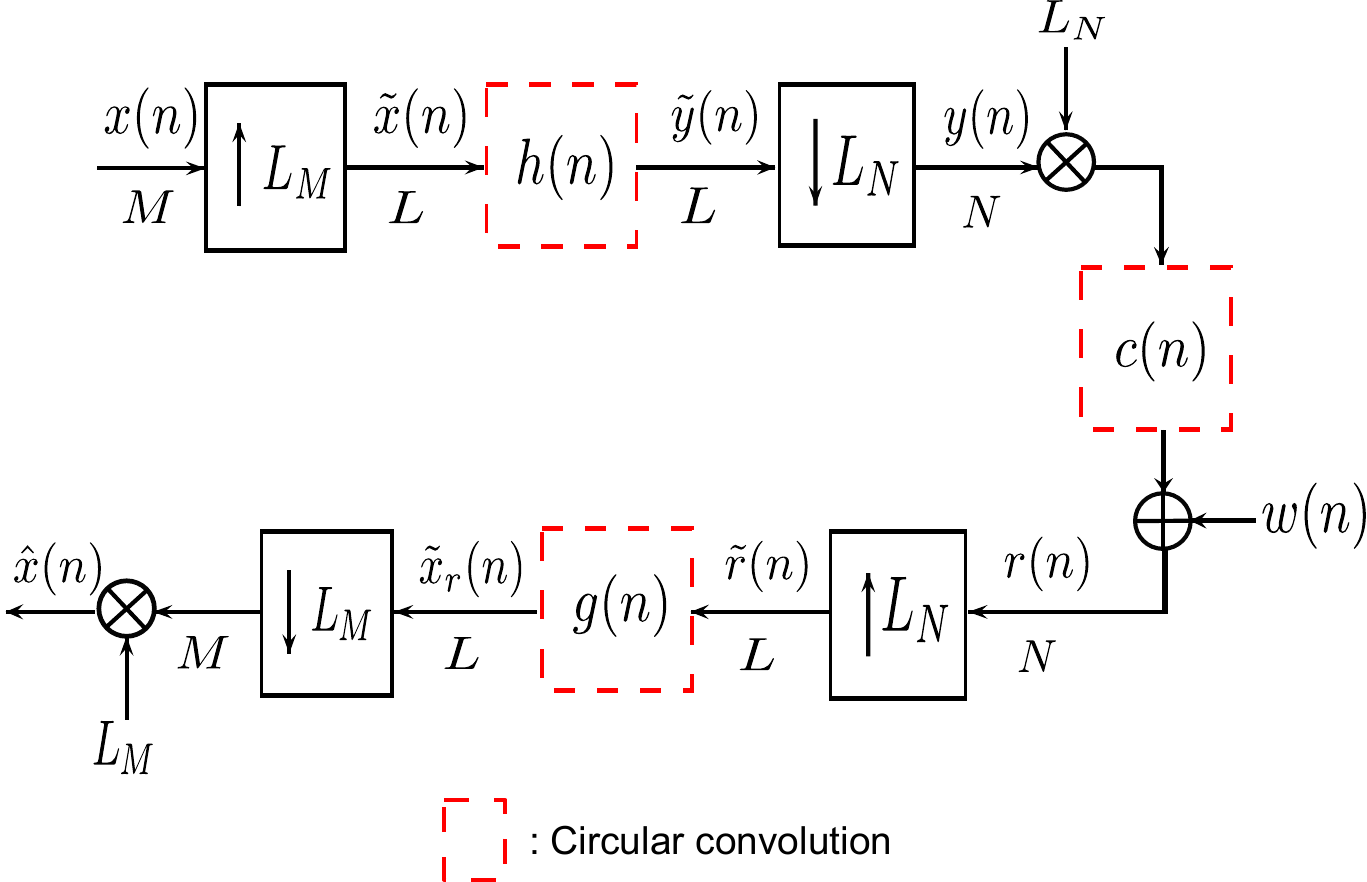}
\caption{SC-FDMA equivalent time domain system model}
\label{fig:figure_up_down2}
\end{figure}
Let us consider the system representation in \rref{fig:figure_up_down} which is obtained by replacing both the transmitter and receiver by their equivalent models in \rref{fig:Tx_model} and \rref{fig:rx_model}. 
In order to develop the time domain model, we use two important frequency-time equivalences. On the one hand, it is well established that a frequency domain multiplication with a frequency response $H$ of length $L$ translates into circular convolution in the time domain with the time domain filter $h(n) = IFFT_L(H)$ which writes as follows: 
\beq 
h(n)= \frac{1}{L} \sum_{p=0}^{L-1}H_p \Omega_L^{p n}
\eeq 
The output symbols $\tilde{y}(n)$ of the circular convolution of filter $h(n)$ and symbols $\tilde{x}(n)$ can be expressed as follows: \\
\beq 
\tilde{y}(n) = \sum_{m=0}^{L} \tilde{x}(m) h(<n-m>_L) = h(n)\circledast \tilde{x}(n)
\eeq
where $<.>_L$ denotes the modulo $L$ operator and $\circledast$ stands for circular convolution. \\ In a similar way, by defining 
$g(n)= \frac{1}{L} \sum_{p=0}^{L-1}G_p \Omega_L^{p n}$
The received equalized symbols $\tilde{x}_r(n)$ are expressed as follows: 
\beq 
\tilde{x}_r(n) = \sum_{m=0}^{L} \tilde{r}(m) g(<n-m>_L) = g(n)\circledast \tilde{r}(n)
\eeq
On the other hand, inserting a CP -of length $N_g$ longer than the channel memory- at the output of the transmitter and removing it at the input of the receiver, yields a circular time domain convolution. As a consequence, using the two aforementioned properties, blocks delimited with dashed lines in  \rref{fig:figure_up_down} are equivalent to circular convolution in the time domain as represented in \rref{fig:figure_up_down2}. 

\subsection{\textbf{Spectral shaping SC-FDMA as a circular convolution}}\label{sec:spectrumshaping}
\begin{figure}[t]
\centering
\includegraphics[width=3.5in]{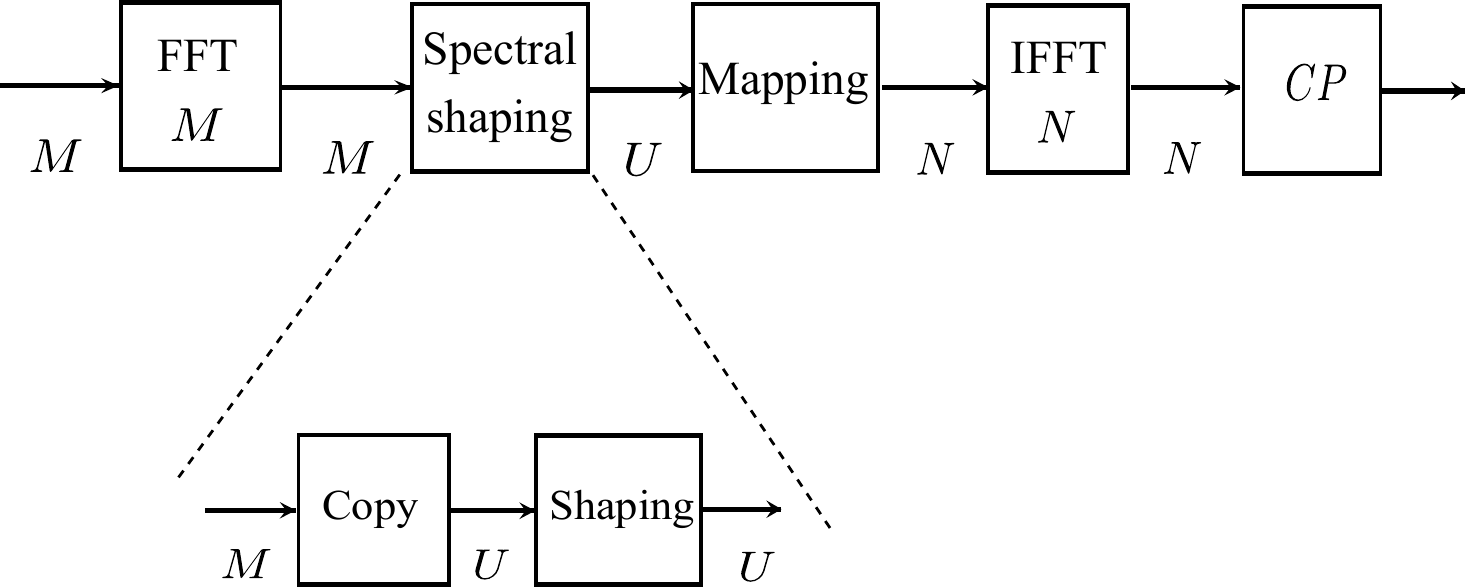}
\caption{Spectral shaping scheme}
\label{fig:spec_shaping}
\end{figure}

\begin{figure}[t]
\centering
\includegraphics[width=2in]{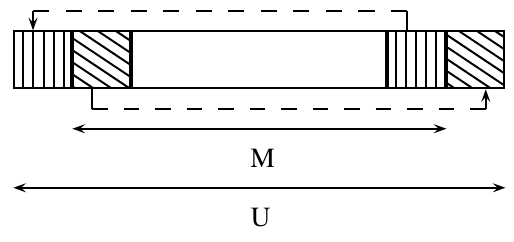}
\caption{Copying from a length $M$ to $U \leq 2M$}
\label{fig:Copying}
\end{figure}
\begin{figure}[!t]
\centering
\includegraphics[width=3.5in]{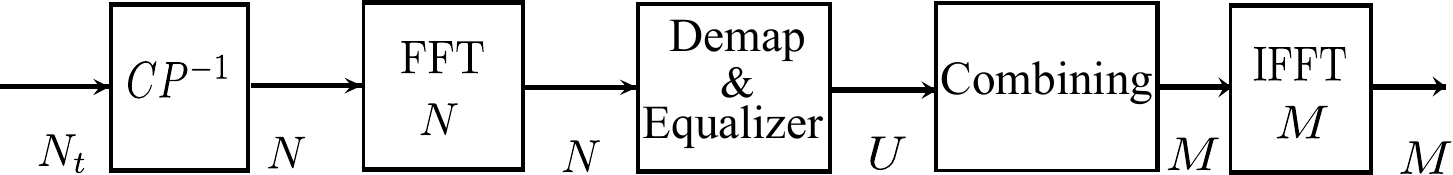}
\caption{Receiver structure of SC-FDMA using spectral shaping of length $U \geq M$}
\label{fig:rx_ss}
\end{figure}
Spectral Shaping (SS) or frequency domain precoding consists of multiplying the $M$-FFT outputs by a shaping window which leads to lower side-lobes and thus a reduced PAPR. \rref{fig:spec_shaping} depicts the Spectrally Shaped SC-FDMA (SS-FDMA) scheme. Spectral shaping is a frequency domain processing inserted between the $M$-FFT and $N$-IFFT. 
If the length $U$ of the spectrum shape is not larger than $M$, then the shaping will only consist of an element wise multiplication with the shaping window. However, if the length of the window shape exceeds the number of $M$-FFT outputs (i.e. $U > M$), then the number of frequency symbols needs to be increased by the so-called copying block \cite{Kawamura_iscws_2008}. A copy consists of appending both a cyclic prefix and suffix to the $M$-FFT symbol until the desired length $U$ is reached as depicted in \rref{fig:Copying}. \\
However, the copying (or duplication) has been proposed in the case of root raised cosine shaping, the length of which would not exceed $2M$. Yet, this process can be extended to a more general scheme which consists of the aforementioned repetition block illustrated in \rref{fig:up_sampl} allowing for a length of spectral shaping up to $U \leq L$. As such, the general model depicted in \rref{fig:figure_up_down} also applies for the general spectral shaping of length $U$ up to $L$. The spectral shaping window can thus be included in the frequency response of the transmit filter $H$.  \\
An additional difference between SS-FDMA and classical SC-FDMA lies in the receiver processing(see \rref{fig:rx_ss}). Indeed, since some of the $M$ symbols have been duplicated to reach a length $U \geq M$ a "frequency combining" block has to be added before passing through the final $M$-IFFT in order to combine received symbols issuing from the duplicated transmitted symbols. The frequency combining as presented in \cite{Kawamura_iscws_2008} is in fact a special case of the stacking operation in (\ref{equ:stacking}). Note that since the stacking at the receiver has parameters $(L,M)$ and the stacking inputs contain $U \geq M$ non zeros elements, the operation is no longer transparent (i.e. no longer a simple multiplication with $\frac{1}{L_M}$). This impacts channel equalization as will be discussed in section \ref{sec:lte}.  
In a nutshell, the spectral shaping scheme is also covered by the general model proposed in \rref{fig:figure_up_down}. 
Next section presents a summary of different system parameters of the general model depicted in \rref{fig:figure_up_down} allowing to find the special cases of both classical FDMA and SS-FDMA with different spectrum shapes.  
 
\subsection{\textbf{Special cases of the general scheme: LFDMA and SS-FDMA}}
\subsubsection{3GPP Localised SC-FDMA}\label{sec:lfdma_filter}
In the localised mapping of the 3GPP proposed SC-FDMA, the mapping frequency response $H $ consists of a block of $M$ non-zero frequency bins out of $L$. 
It can thus be viewed as a rectangular shaping in the frequency domain with length $M$ which satisfies:   
\beq
H_k  = 
\left\{ \begin{array}{ll}
      	 1 & if \: \:  0 \leq k \leq M-1 \\  
         0 & if \: \:  M \leq k \leq L-1
\end{array} \right.
\label{equ:transmit_filter}
\eeq
In an equivalent way, the LFDMA equalizer and demapper $G$ consists of a frequency response with only $M$ non zero elements i.e. $|G_p| = 0$ if $p \geq M$ as depicted in \rref{fig:demapping_model}.

\subsubsection{Raised cosine Spectrally Shaped LFDMA}\label{sec:ssfdma_filter}
\begin{figure}[t]
\centering
\includegraphics[width=3in]{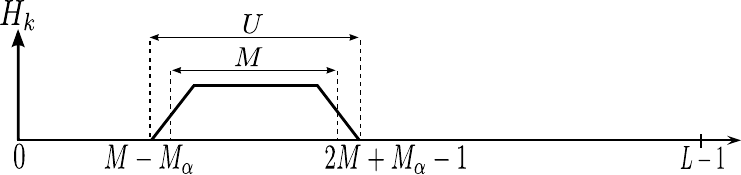}
\caption{Spectral shaping filter with root raised cosine}
\label{fig:rcf_ss}
\end{figure}
Let us consider a root raised cosine spectral shaping with a roll-off factor $\alpha$. Let $M_{\alpha} = \lfloor \alpha \frac{M}{2} \rfloor$ where $\lfloor . \rfloor$ denotes the floor operator. The length of the root raised cosine window $U$ satisfies $ 0 \leq U = M + 2 M_{\alpha} \leq 2M$. To avoid border overlapping effects for the fractional case, the user is placed in the $2^{nd}$ block of $M$ resource frequencies. More precisely, the spectral shaping window depicted in \rref{fig:rcf_ss} writes as follows: 
\beqas 
H_k \left\{
\begin{array}{ll} 
   \neq 0 & if  \: k = \{M-M_{\alpha}, \hdots, 2M+M_{\alpha}-1\}\nnb \\ 
    = 0 & else  
 \end{array} \right. 
\eeqas 
The equalizer frequency response consists of $U$ non zero frequency bins located at indexes $\{M-M_{\alpha}, \hdots, 2M+M_{\alpha}-1\}$. 
Next sections provide analytical expression of both PSD and SINR for the general SC-FDMA scheme where frequency responses of the transmit window $H$, the channel $C$ and the equalizer $G$ are assumed general i.e. no restriction on the number of non zero elements is made. Numerical applications are then presented for the two special cases of LFDMA and raised cosine SS-FDMA. \\
\section{Power Spectral Density of general SC-FDMA}\label{sec:psd}
\begin{figure}[!t]
\centering
\includegraphics[width=3in]{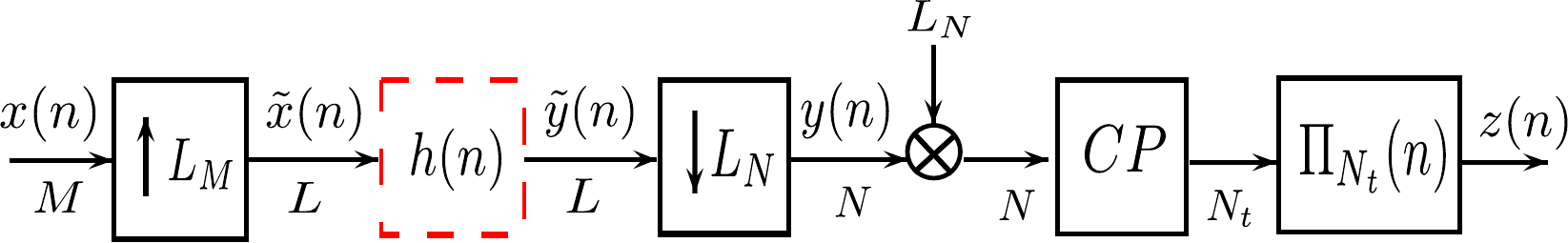}
\caption{Transmitter model with pulse shaping}
\label{fig:tx_pulse_shape}
\end{figure}
The power spectral density analysis is of paramount importance for system design since it allows to assert that the transmitter respects the spectrum or transmission mask usually defined to limit the inter-channel interference. In the case of multi-user communications, it is also valuable to multiplex users respecting some allowed inter-users interference through resource allocation \cite{babar_2011}. \\
Consider the general scheme in \rref{fig:tx_pulse_shape}. A CP is appended to the samples $L_N y(n)$ to form the LFDMA symbols of length $N_t$. A pulse shaping waveform $\Pi_{N_t}(t)$ of length $N_t$ symbols is used at the front end of the transmitter. \\
The transmitted LFDMA symbols $z(n)$ write as follows:  
\beq 
z(n) =   L_N  \sum_{l=-\infty}^{\infty} y(<n-lN_g>_N)\Pi_{N_t}(n-lN_t) \nnb \\
\eeq
Using the modulo arithmetic equality $<n-lN_g>_N L_N = <(n-lN_g)L_N>_L$, we can write the transmitted symbols as:  \\
\beqa 
z(n) &= &  L_N  \sum_{l=-\infty}^{\infty}\sum_{p=0}^{L-1} \tilde{x}_{p,l} h(<(n- lN_g)L_N-p>_L) \Pi_{N_t}(n-lN_t) \nnb \\
& =&  L_N  \sum_{l=-\infty}^{\infty}\sum_{p=0}^{M-1} x_{p,l} h(<(n- lN_g)L_N -p L_M >_L)
 \Pi_{N_t}(n-lN_t)  \\ \nnb  
\eeqa

The autocorrelation $R_{z}(n,m)$ of symbols $z(n)$ can be derived as follows :  \\
\beq  
\begin{multlined}
R_{z}(n,m)  =  E\left[ z(n)z^*(n-m)\right]  \\
\shoveleft  = L_N^2 \sum_{l=-\infty}^{\infty}\sum_{p=0}^{M-1}\sum_{l'=-\infty}^{\infty}\sum_{p'=0}^{M-1}E\left[ x_{p,l}x_{p',l'}^*\right]
 \vphantom{\sum_{l=-\infty}^{\infty}}h^*(<(n-m-lN_g)L_N-p'L_M>_L)\Pi_{N_t}(n-lN_t)\\ 
 \shoveleft[0.5cm] h(<(n-lN_g)L_N-pL_M>_L)  \Pi_{N_t}(n-m-l'N_t)  \\ \nnb
= L_N^2 \sigma_x^2 \sum_{l=-\infty}^{\infty}\sum_{p=0}^{M-1}h(<(n-lN_g)L_N -pL_M>_L)  h^*(<(n-m-lN_g)L_N - pL_M>_L) \\
\shoveleft[0.5cm]   \Pi_{N_t}(n-lN_t) \Pi_{N_t}(n-m-lN_t) \\
\end{multlined}
\label{equ:autocorr}
\eeq %
%
In order to further simplify the above expression one needs to compute $\sum_{p=0}^{M-1}h(<nL_N -pL_M >_L)h^*(<(n-m)L_N-pL_M>_L)$. When writing the filter $h(n)$ in the frequency domain, the expression nicely simplifies as shown in the following:   \\%
\beq
\begin{multlined}
\sum_{p=0}^{M-1}h(<nL_N -pL_M >_L)h^*(<(n-m)L_N-pL_M>_L)  \\
\shoveleft = \frac{1}{L^2} \sum_{k=0}^{L-1}\sum_{k'=0}^{L-1}H_k H_{k'}^* \left(\sum_{p=0}^{M-1} \Omega_M^{p(k'-k)}\right)\Omega_L^{(kn -k'(n-m))L_N}   \\
\shoveleft = \frac{1}{L^2} \sum_{s=0}^{L_M-1}\sum_{s'=0}^{L_M-1}\sum_{r=0}^{M-1}\sum_{r'=0}^{M-1}H_{sM+r} H_{s'M+r'}^* 
 \left(\sum_{p=0}^{M-1}\Omega_M^{p(r'-r)}\right) \Omega_L^{(n(sM+r) -(n-m)(s'M+r'))L_N}   \\ \nnb
\shoveleft = \frac{M}{L^2}  \sum_{r=0}^{M-1}  \sum_{s=0}^{L_M-1}H_{sM+r} \Omega_{L_M}^{snL_N} \sum_{s'=0}^{L_M-1} H_{s'M+r}^* \Omega_{L_M}^{s'(n-m)L_N}\Omega_N^{rm}  \nnb  \\
\shoveleft = \frac{M}{L^2} \sum_{r=0}^{M-1} h_{r}(<n>_{L_M}) h_r^*(<n-m>_{L_M}) \Omega_{N}^{rm} \nnb \\
\end{multlined}
\eeq  
 
\noindent where  $h_{r}(<n>_{L_M}) = \sum_{s=0}^{L_M-1}H_{sM+r} \Omega_{L_M}^{snL_N}$. It should be noted that this function $h_{r}(<n>_{L_M})$ is $L_M$ periodic and satisfies  $ h_{r}(<n+N>_{L_M})= h_r(<n>_{L_M})$ since $N$ is a multiple of $L_M$.\\
The transition from the second to the third equality is based on the euclidean division of $k$ and $k'$ over $M$ leading to $k = sM+r$ and $k' = s'M+r'$ where $r, \: r' \in{0, \hdots, M-1}$ and $s, \: s' \in {0, \hdots, L_M-1}$. 
The following exponential identity has also been used: \\
\beq
\sum_{p=0}^{M-1}\Omega_M^{p r}= 
\left\{
\begin{array}{ll}
M & if \: r = kM \\
0 & else \\
\end{array}\right. 
\eeq
Equation (\ref{equ:autocorr}) can be finally written as follows:  
\beq
\begin{multlined}
\shoveleft R_{z}(n,m) =\frac{M  \sigma_x^2 }{N^2}  \sum_{l=-\infty}^{\infty} \sum_{r=0}^{M-1} h_{r}(<n-lN_g>_{L_M})\Omega_N^{r m}  \\ 
h_r^*(<n-m-lN_g>_{L_M})  \Pi_{N_t}(n-lN_t) \Pi_{N_t}(n-m-lN_t)\\ \nnb
\end{multlined}
\eeq 
It can be noticed that $R_{z}(n,m) =R_{z}(n+N_t,m)$, thus $R_{z}$ is $N_t$-periodic in time. This allows to derive a stationary autocorrelation of symbols $z(n)$ by averaging over the time domain dimension $n$ as follows:  
\beq  
\begin{multlined}
\overline{R}_{z}(m) = \frac{1}{N_t}\sum_{n=0}^{N_t-1} R_{z}(n,m) \nnb \\
\shoveleft = \frac{M  \sigma_x^2 }{N_t N^2} \sum_{r=0}^{M-1} \sum_{n=0}^{N_t-1}  \sum_{l=-\infty}^{\infty}  h_r^*(<n-m-lN_g>_{L_M})	h_{r}(<n-lN_g>_{L_M}) \Omega_N^{rm} \\ 
\shoveleft \Pi_{N_t}(n-lN_t) \Pi_{N_t}(n-m-lN_t) \nnb \\
\shoveleft =   \frac{M  \sigma_x^2 }{N_t N^2}  \sum_{r=0}^{M-1}\sum_{n=-\infty}^{\infty} h_r(<n>_{L_M}) h_r^*(<n-m>_{L_M}) \Pi_{N_t}(n) \Pi_{N_t}(n-m)\Omega_N^{rm} \nnb \\
\shoveleft =   \frac{M  \sigma_x^2 }{N_t N^2}\sum_{r=0}^{M-1}\sum_{n=-\infty}^{\infty} \tilde{h}_r(n) \tilde{h}_r^*(n-m)\Omega_N^{rm} \nnb \\
\shoveleft[0.5 cm] =    \frac{M  \sigma_x^2 }{N_t N^2}\sum_{r=0}^{M-1} R_{\tilde{h}_r}(m) \Omega_N^{r m} \\
\end{multlined}
\eeq 
where we define the equivalent transmit filter $\tilde{h}_r$ as:  \\
\beqs 
\tilde{h}_r(n) =  h_{r}(<n>_{L_M}) \Pi_{N_t}(n) = \sum_{s=0}^{L_M-1}H_{sM+r} \Omega_{L_M}^{snL_N}\Pi_{N_t}(n)
\eeqs
 and the autocorrelation function of $\tilde{h}_r$ is defined by: $R_{\tilde{h}_r}(m) = \sum_{n=-\infty}^{\infty} \tilde{h}_r(n) \tilde{h}_r^*(n-m)$.\\
As a consequence, the power spectral density of LFDMA can be written as follows:  
\beqa 
\overline{S_{z}}(f) &=& \sum_{m=-\infty}^{\infty}\overline{R}_{z}(m) e^{-2j \pi mf} \nnb \\
&=& \frac{M  \sigma_x^2 }{N_t N^2}\sum_{r=0}^{M-1} \sum_{m=-\infty}^{\infty} R_{\tilde{h}_r}(m) \Omega_N^{r m} e^{-2j \pi mf} \nnb \\
&=& \frac{M  \sigma_x^2 }{N_t N^2}\sum_{r=0}^{M-1} \left |\tilde{H}_r(f-\frac{r}{N})\right |^2 \nnb \\
&=& \frac{M  \sigma_x^2 }{N_t N^2}\sum_{r=0}^{M-1} \left |\sum_{s=0}^{L_M-1}H_{sM+r} \Psi_{N_t} (f-\frac{sM + r  }{N})\right |^2  \nnb \\
\eeqa 
where  $\Psi_{N_t}(f)$ is the Energy Spectral Density (ESD) of the time domain shaping filter $\Pi_{N_t}(n)$. \\ 
It is interesting to point out that the final PSD expression can be interpreted in the sense that due to the repetition block, each symbol out of the $M$ LFDMA input symbols undergoes a global frequency response which is the sum of all $M$-evenly spaced frequency responses $H_{sM+r}$ where $s \in [0:L_M-1]$. \\
\section{Applications of PSD of localised FDMA with general spectral shaping}\label{sec:psd_appli}
\subsection{\textbf{PSD Rectangular shaping: LTE LFDMA}}
As previously explained in section \ref{sec:timedomain}, the equivalent transmit window in LFDMA implementation is a rectangular window of length $M$ placed in frequency bins ${0, \hdots, M-1}$. 
It follows that  
\beq 
\overline{S}_{z}(f) = \frac{M  \sigma_x^2 }{N_t N^2}\sum_{r=0}^{M-1} \left |\Psi_{N_t}(f-\frac{r}{N})\right |^2  \nnb \\
\eeq
A (digital) sampled rectangular filter of length $N_t$ has a dirichlet kernel transfer function described as follows \cite{Waterschoot_2010}:  \\
\beq  
\Psi_{N_t}(w) = sinc_{N_t} (w) =
\left\{ \begin{array}{ll}
-1^{w(N_t-1)} & if \: \: w \in \mathbb{Z} \\
\frac{sin(N_t w/2 )}{N_t sin(w/2)} & otherwise
\end{array} \right.
\eeq 
\begin{table}
\begin{minipage}[c]{0.45\textwidth}%
\includegraphics[width=3in]{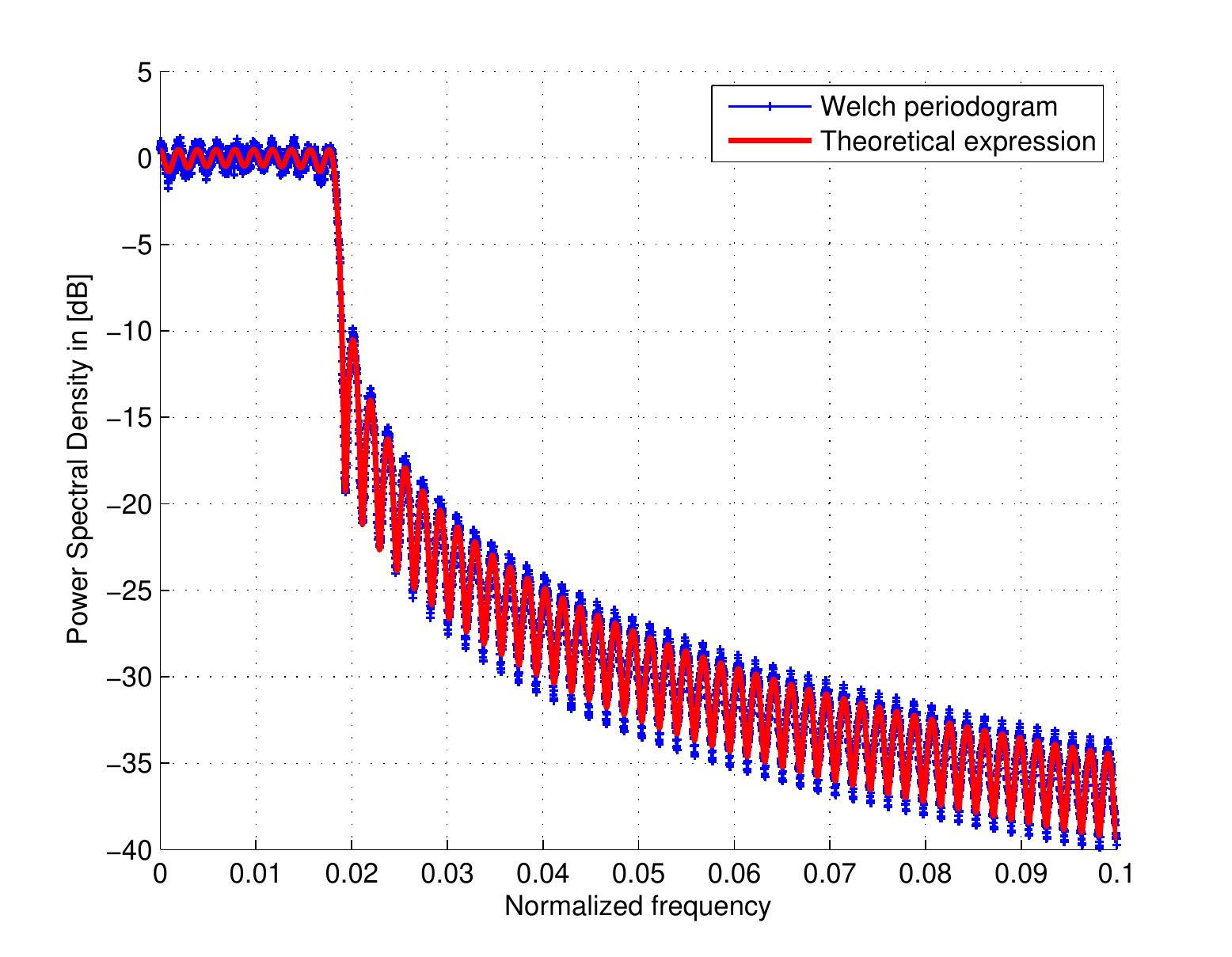}
 \captionof{figure}{Rectangular shaping with $M=10$, \\ $N=512$, and $CP=31$}
\label{fig:PSD_rof0_frac}
\end{minipage} 
\begin{minipage}[c]{0.45\textwidth}%
\includegraphics[width=3in]{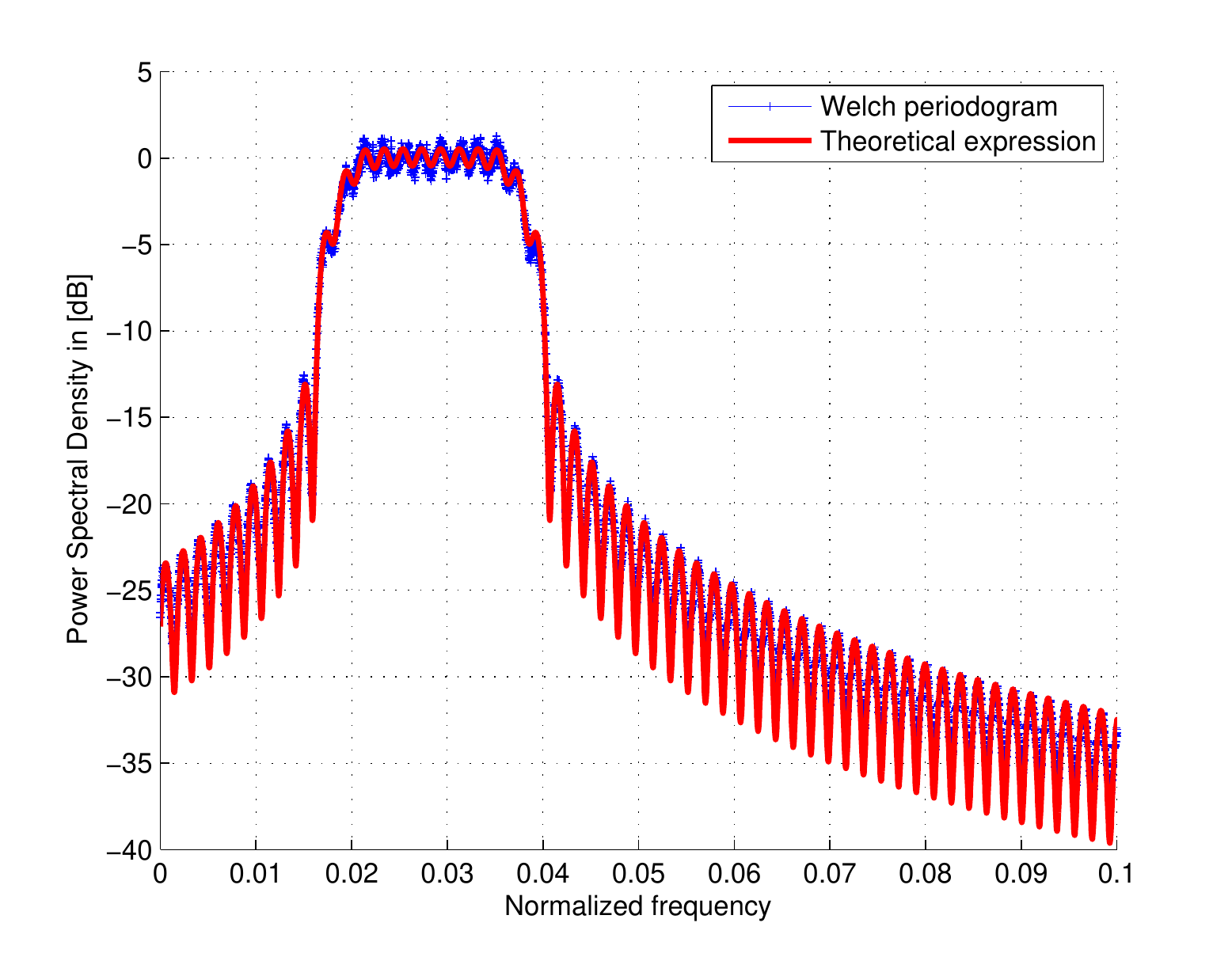}
\captionof{figure}{SRC shaping with $M=10$, $N=512$, \\ roll-off$=0.35$ and $CP=31$}
\label{fig:PSD_rof035_frac}
 \end{minipage}
\end{table}
\rref{fig:PSD_rof0_frac} plots the power spectral density of rectangular spectrally shaped LFDMA obtained by a Welch periodogram with $50 \% $ overlapping for the fractional rate ($M=10$, $N=512$) with CP length $N_g = 31$. The theoretical and estimated PSD are perfectly matched. 
\subsection{\textbf{PSD with root raised cosine spectral shaping}} 

For the SS-FDMA with root raised cosine, the transmit window $H$ is expressed in section \ref{sec:timedomain}.  \\
The power spectral density reads as follows: 
\beq
\overline{S}_{z}(f) = \frac{M  \sigma_x^2 }{N_t N^2}\sum_{r=0}^{M-1} \left |\Gamma_{N_t}^{(r)}(f)\right |^2  \nnb \\
\eeq
where $\Gamma_{N_t}^{(r)}$ is the equivalent transmit response given in equ.(\ref{equ:freq_resp_rcf}).
\beqa
\Gamma_{N_t}^{(r)}(f)= \left\{
\begin{array}{ll}
 	 H_{M+r}   \Psi_{N_t}(f- \frac{M + r}{N}) + H_{2M+r}  \Psi_{N_t}(f- \frac{2M+r}{N}) & if  \: r = {0,\hdots,M_{\alpha}-1} \\
  	 H_{M+r}   \Psi_{N_t}(f- \frac{M + r}{N})  &  if  \: r = {M_{\alpha}, \hdots, M-M_{\alpha}-1} \\
      H_{M+r}   \Psi_{N_t}(f- \frac{M + r}{N}) + H_{r} \Psi_{N_t}(f-\frac{r}{N}) & if  \: r = {M-M_{\alpha},\hdots, M-1}  
 \end{array} \right. 
 \label{equ:freq_resp_rcf}
\eeqa
\rref{fig:PSD_rof035_frac} plots the power spectral density of root raised cosine spectrally shaped LFDMA obtained by a Welch periodogram with $50 \% $ overlapping for the fractional rate ($M=10$, $N=512$) with CP length $N_g = 31$. Both simulations and theoretical expressions match.  \\
\section{Localised SC-FDMA SINR}\label{sec:sinr}
\begin{figure}[!t]
\centering
\includegraphics[width=3in]{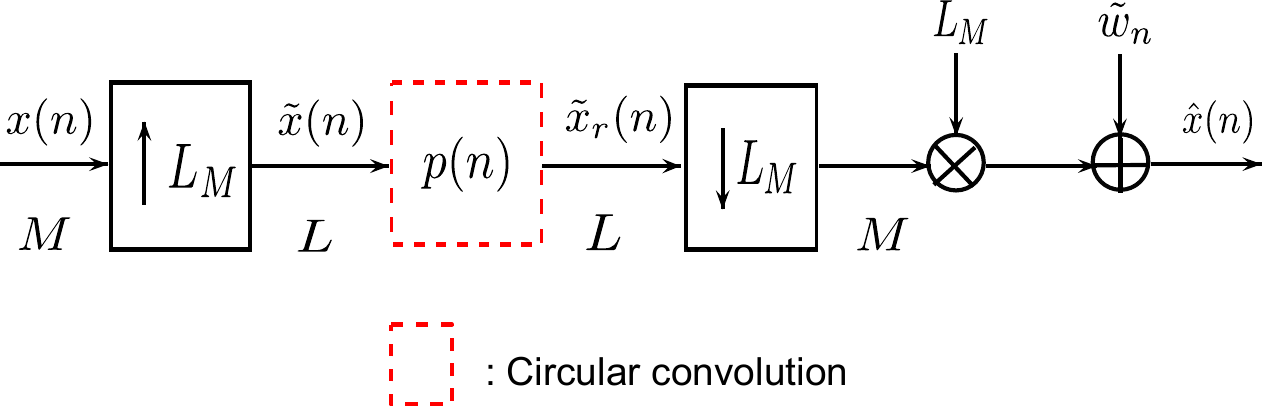}
\caption{SC-FDMA simplified system model}
\label{fig:figure_up_down3}
\end{figure}
The system depicted in \rref{fig:figure_up_down2} can be further simplified when considering an up-sampled version of the channel frequency response. In other words, we define the expanded frequency response $\tilde{C}_k$ obtained as follows:  
\beq
\tilde{C}_{k}  = \left \{
\begin{array}{ll}
C_k & 0 \leq k \leq N-1 \\ 
0  & N \leq k \leq L 
\end{array}
\right.
\eeq
Thus $\tilde{c} = IFFT \left( \tilde{C} \right) $ is the up-sampled IFFT of the frequency response $C_k$. 
This allows for a compact system model as depicted in  \rref{fig:figure_up_down3} where $p(n)$ is the over-all system time response which writes as  
\beq 
p(n) = h(n) \circledast \tilde{c}(n)\circledast g(n) \hspace{10 mm} 0 \leq n \leq L-1
\eeq 
The frequency response of the overall system is $P_k = H_k\tilde{C}_k G_k$ for $k \in [0:L-1]$. 
\begin{figure}[t]
\centering
\includegraphics[width=2.5in]{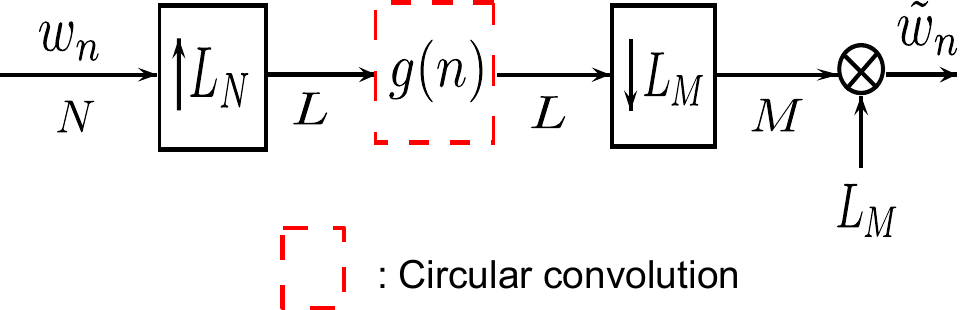}
\caption{Equivalent noise}
\label{fig:figure_noise_interm}
\end{figure}
The equivalent noise $\tilde{w}(n)$ is obtained by up-sampling the additive noise $w(n)$ by a factor $L_N$, circularly convolving with the equalizer $g(n)$ and down-sampling it by a factor $L_M$ as in \rref{fig:figure_noise_interm}. The final multiplication factor in the general scheme arises from the $L_M$ downsampling-stacking inversion.  
Let us consider the general model depicted in figure \rref{fig:figure_up_down3}. For a block of $N$ received symbols, the demapped equalized symbols $\hat{x}(n)$ write as follows, $\forall n \in [0:M-1]$ :  
\beqa 
\hat{x}(n) &  =& L_M \tilde{x}_r(n L_M) + \tilde{w}(n)\nnb \\
		   &=& L_M \left(\sum_{k=0}^{L-1} p(<n L_M -k>_L) \tilde{x}(k)\right)+ \tilde{w}(n)  \nnb \\
		   &=&L_M \left( \sum_{m=0}^{M-1} p(<(n-m)L_M>_L) \tilde{x}_r(m L_M) \right) + \tilde{w}(n) \nnb \\  
   		   &=&L_M \left( \sum_{m=0}^{M-1} p(<(n-m)L_M>_L) x(m)  \right)+ \tilde{w}(n)  \nnb \\
&=& x_u(n) + x_i(n) + \tilde{w}(n) \nnb 
		   \label{equ:estimated}
\eeqa 
where we define the: \\
\begin{itemize}
\item  useful signal $x_u(n)=  L_M p(0)x(n)$ 
\item  interfering signal $x_i(n) = L_M \sum_{n \neq m}  p(<(n-m)L_M>_L) x(m)$
\item  equivalent noise $\tilde{w}(n)$.
\end{itemize}

SINR is a valuable metric that relates the power of the desired signal at the receiver to the amount of interference and noise. 
It is thus defined as follows: 
\beq 
SINR =\frac{P_u}{\sigma_i^2 + \sigma_{\tilde{w}}^2}
\eeq 
where  $P_u = E[|x_u(n)|^2]$, $\sigma_i^2 = E[|x_i(n)|^2]$, $\sigma_{\tilde{w}}^2= E[|\tilde{w}(n)|^2]$ are the powers of the useful resp. interfering and noise terms.\\

\subsection{\textbf{The useful term power $P_u$ }}

The useful term power is $P_u =( L_M)^2|p(0)|^2 \sigma_x^2$ where $\sigma_x^2$ is the variance of the transmitted symbols $x_n$. From the FFT definition, $p(0) = \frac{1}{L} \sum_{k=0}^{L-1}P_k$. \\
Thus, the useful power can be written as :  
\beq 
P_u = \frac{\sigma_x^2}{M^2} \left|\sum_{k=0}^{L-1}P_k\right|^2 
\label{equ:useful_var}
\eeq 
\subsection{\textbf{The interfering term power $\sigma_i^2$ }} 
For $n \in [0:M-1] \hspace{5 mm}$ the received symbols $ x_r(n)= x_u(n) + x_i(n) = L_M \sum_{m=0}^{M-1} p(<(n-m)L_M>_L) x(m) $. The power of the received symbols $x_r(n)$ writes as: 
\beq
\sigma_r^2  =  E[| x_r(n)|]^2 = \left(L_M\right)^2\sigma_x^2  \sum_{m=0}^{M-1}\left| p(<(n-m)L_M>_L)\right|^2 =  P_u + \sigma_i^2   
\label{equ_powe_rx}
\eeq
The received signal is the result of up-sampling and down-sampling the original stationary symbols with the same factor $L_M$. Thus, it is a stationary process and its power is not dependent on the time index $n$. More specifically, for $n \in [0:M-1]$, let $ \tilde{p}(n) = p(nL_M)$, i.e. $\tilde{p}$ results from down-sampling the global filter $p$ by a factor $L_M$. \\
From the previous result of down-sampling in \rref{fig:down_sampl}, $\tilde{p}(n) = IFFT_M(P_{L,M})$  where $P_{L,M}(r) = \frac{M}{L} \sum_{l=0}^{L_M-1} P_{lM + r}$ is the stacking output for $r \in [0:M-1]$.\\
Thus, $p(nL_M)$ can be written as :  
\beq 
p(nL_M) =\tilde{p}(n) = \sum_{r=0}^{M-1}P_{L,M}(r)\Omega_M^{kr}  
\eeq  
From Parseval identity, we have: 
\beq
\sum_{n=0}^{M-1} \left| p(nL_M) \right|^2 = \frac{1}{M}\sum_{r=0}^{M-1}\left| P_{L,M}(r) \right|^2 = \frac{1}{M}\left(\frac{M}{L}\right)^2\sum_{r=0}^{M-1}\left| \sum_{l=0}^{L_M-1} P_{lM + r}\right|^2 
\eeq 
It follows that 
\beq
\sigma_r^2 = \frac{\sigma_x^2}{M} \sum_{r=0}^{M-1}\left| \sum_{l=0}^{L_M-1} P_{lM + r}\right|^2 \\
\eeq 
Inserting this result in equ.(\ref{equ_powe_rx}), the power of the interfering term is as follows:  
\beq  
\sigma_i^2    = \sigma_r^2 - P_u = \frac{\sigma_x^2 }{M^2}\left(M\sum_{r=0}^{M-1}\left| \sum_{l=0}^{L_M-1} P_{lM + r}\right| ^2 - \left|\sum_{k=0}^{L-1}P_k\right|^2\right)\nnb\\
		   \label{equ:interf_var}
\eeq  
\subsection{\textbf{The noise power $\sigma_{\tilde{w}}^2$ }} 

Let us consider the equivalent noise depicted in \rref{fig:figure_noise_interm}. Due to up-sampling by a factor $L_N$ and down-sampling by a different factor $L_M$, the noise is not necessarily stationary. What can be shown is that unless special conditions are imposed on the equalizing filter $g(n)$, the noise is cyclo-stationary. This means that the SINR will have instantaneous values depending on the position of the noise sample in the block. 
However, we can derive a \textit{mean} SINR by stationarizing the noise process leading to a noise variance (refer to appendix):  
\beq 
\sigma_{\tilde{w}}^2 = E[|\tilde{w}(n)|^2] =  \frac{\sigma_w^2 N}{M^2} \sum_{r=0}^{N-1}\left|G_k \right|^2
\label{equ:noise_var}
\eeq 
Merging the three results equ.(\ref{equ:useful_var}), equ.(\ref{equ:interf_var}) and equ.(\ref{equ:noise_var}) the mean SINR reads as in equ.(\ref{equ:SINR_1}). 
  \beq  
SINR = \frac{\left|\sum_{k=0}^{L-1}P_k\right|^2}{M\sum_{r=0}^{M-1}\left| \sum_{l=0}^{L_M-1} P_{lM + r}\right|^2 - \left|\sum_{k=0}^{L-1}P_k\right|^2 + N \frac{\sigma_w^2}{\sigma_x^2} \sum_{k=0}^{L-1}\left|G_k\right|^2 }
\label{equ:SINR_1}
\eeq   
\subsection{\textbf{SINR function of SNR}}
It is interesting to have a formulation of the SINR in terms of SNR, to have an easy asymptotic interpretation of the system performance function of SNR. 
The SNR is defined as the ratio of the signal power at the input of the receiver to the noise power. \\
The signal power computed at the input of the receiver is $P = \sigma_x^2 \frac{M}{N^2} \sum_{k=0}^{L-1}\left|H_k \tilde{C}_k\right|^2$. \\
The noise power at the input of the receiver is function of the sampling rate and reads: $ \sigma_b^2 = N_0 F_s$ where $F_s = \frac{N}{M} R_s$ is the sampling rate at the input of the receiver and $R_s$ is the user sampling rate. It should be noted that to fulfil the Shannon sampling theorem for rectangular LFDMA ($F_s \geq 2 R_s$), $N$ should satisfy $N \geq 2 M$. In the case of raised cosine spectral shaping, $N$ should satisfy $N \geq 2(1+\alpha)M$.\\ 
It follows that  \beq 
\frac{E_s}{N_0} = \frac{P T_s}{N_0} =  \frac{P}{N_0 R_s} =  \frac{\sigma_x^2}{\sigma_w^2}\frac{1}{N}\sum_{k=0}^{L-1}\left|H_k \tilde{C}_k \right|^2 
\eeq 
Thus, using the following result 
\beq 
 N \frac{\sigma_w^2}{\sigma_x^2} = \left(\frac{E_s}{N_0}\right)^{-1} \sum_{k=0}^{L-1}\left|H_k \tilde{C}_k\right|^2 
 \label{equ:SNR}
\eeq 
Equ.(\ref{equ:SINR_1}) leads to the SINR in equ.(\ref{equ:SINR_2}).

 \beq  
SINR = \frac{\left|\sum_{k=0}^{L-1}P_k\right|^2}{M\sum_{r=0}^{M-1}\left| \sum_{l=0}^{L_M-1} P_{lM + r}\right|^2 - \left|\sum_{k=0}^{L-1}P_k\right|^2 + \left(\frac{E_s}{N_0}\right)^{-1} \sum_{k=0}^{L-1}\left|H_k \tilde{C}_k\right|^2 \sum_{k=0}^{L-1}\left|G_k \right|^2}
 \label{equ:SINR_2}
\eeq 
 It should be noted that unlike SINR expressions in \cite{sinr_3GPP}, the above SINR analytical expressions apply to the fractional case as well.


In the next section we will apply results of the SINR to some SC-FDMA implementations with two implementations of linear equalizers, namely Zero Forcing (ZF) and Minimum Mean Square Error (MMSE) equalizers.   \\
\subsection{\textbf{Linear equalizers: MMSE and ZF}}
\subsubsection{Zero Forcing (ZF) equalizer}
Let us derive expressions of linear equalizers for LFDMA general scheme beginning with ZF equalization. The estimated symbols write as follows: 
\beq
\hat{x}_n = L_M \left( \sum_{m=0}^{M-1} \tilde{p}(<n-m>_L) x(m) \right)+ \tilde{w}(n) \nnb \\
\eeq 
where $\tilde{p}$ is the $M$-IFFT of the frequency response $\frac{M}{L}\sum_{s=0}^{L_M -1} P_{sM + r}$.
As such, the frequency response of a ZF equalizer should satisfy  $\sum_{s=0}^{L_M -1} P_{sM + r} = 1$ which yields a solution in the form \cite{Clark_98}: 
\beq
G_k^{ZF} = \frac{H_k^* C_k^*}{\sum_{s=0}^{L_M -1}\left|H_{sM + k}\tilde{C}_{sM + k}\right|^2} 
\eeq
Since $\sum_{s=0}^{L_M -1} P^{ZF}_{sM + r} = 1$, The SINR in equ.(\ref{equ:SINR_2}) simplifies as follows: 
\beq 
SINR^{ZF} = \frac{E_s}{N_0} \frac{M^2}{\sum_{k=0}^{L-1}\left|H_k \tilde{C}_k\right|^2 \sum_{k=0}^{L-1}\left|G_k^{ZF} \right|^2}
\eeq 
where it can be shown that :
\beq 
\sum_{k=0}^{L-1}\left|G_k^{ZF} \right|^2 = \sum_{k=0}^{M-1} \frac{1}{\sum_{s=0}^{L_M-1} \left|H_{k+sM} \tilde{C}_{k+sM} \right|^2}
\eeq 
However, the solution $G_k^{ZF}$ may lead to a large noise enhancement when the channel has zeros in its frequency response. \\ 

\subsubsection{Minimum Mean Square Error (MMSE) equalizer}
As for the MMSE equalizer, the frequency response of the equalizer which minimizes the mean square error $E[|\hat{x}_n - x_n|^2]$ writes as: 
\beq
G_k^{MMSE} = \frac{H_k^* C_k^*}{\sum_{s=0}^{L_M -1}\left|H_{sM + k}\tilde{C}_{sM + k}\right|^2 + \frac{\sigma_W^2}{\sigma_X^2}}
\eeq
where $\sigma_W^2 = N \sigma_w^2$ and $\sigma_X^2 = M \sigma_x^2$ are the variances of the noise $N$-FFT outputs and the symbols $M$-FFT outputs. 
Using equ.(\ref{equ:SNR}), the equalizer writes as a function of SNR as follows:  
\beq 
G_k^{MMSE}= \frac{H_k^* C_k^*}{\sum_{s=0}^{L_M-1}\left|H_{sM + k}\tilde{C}_{sM + k}\right|^2 + \frac{1}{M}  \frac{N_0}{E_s} \sum_{k=0}^{L-1} \left|H_{k} \tilde{C}_{k} \right|^2}
\eeq 

\section{Applications to the SINR of SC-FDMA schemes}\label{sec:lte}
\subsection{\textbf{System configuration}}

\begin{table}[!t]
\begin{minipage}[c]{0.45\textwidth}%
\centering
\begin{tabular}{|l |l| }
\hline
   Bandwidth  & $5$ $MHz$  \\
   \hline
   Sub-frame duration & $0.5$ $ms$ \\
   \hline
   LB size & $66.67$ $\mu s$ \\
   \hline
   N (IFFT size) & $512$ \\
   \hline
   CP duration & $31$ \\
   \hline
\end{tabular}
\captionof{table}{LTE system parameters}
\end{minipage}
\begin{minipage}[c]{0.45\textwidth}%
\centering
\begin{tabular}{|c |c |c| }
\hline
  Tap & Relative delay ($ns$)  & Average power ($dB$)  \\
   \hline
   $1$ & $0$ & $0$ \\
   \hline
   $2$ & $130.2$ & $-9.24$ \\
   \hline
   $4$ & $390.6$ & $-22.8$ \\
   \hline
\end{tabular}
\captionof{table}{Simplified Pedestrian channel A}
\label{tab:channel_delay} 
\end{minipage}

\end{table}

In order to evaluate the SINR using linear equalizers, we simulate a classical LFDMA transmission scheme as depicted in \rref{fig:figure_up_down3}, where bits are first mapped into Quadrature Phase Shift Keying (QPSK). The frequency selective channel is a pedestrian channel A \cite{universal_channel_A} at a speed of $3 km/h$ with maximum excess delay of $410 ns$. 
The power delay profile of the pedestrian channel A had to be adapted (simplified) for the system configuration considering a sample duration of $\frac{66.67 \mu s}{512} = 130.2 ns$ as described in \cite{aspects3gpp}. The resulting simplified channel is illustrated in table \ref{tab:channel_delay}. 
QPSK symbols are upsampled by a factor $L_M$ before being circularly convolved with the global filter $p = h \circledast \tilde{c} \circledast g$ where the channel $c$ is a  realisation of the aforementioned channel considered constant over one SC-FDMA block. The equalizer $g$ is either an MMSE or a ZF equalizer.  \\
To illustrate the generality of the results drawn in section \ref{sec:sinr} and for the sake of brevity, we focus on fractional rates implementations of LFDMA and SS-FDMA namely ($N=512$ and $M=10$). 

\subsection{\textbf{SINR of rectangular shaped LTE classical FDMA}}
As previously discussed, the transmitter frequency response $H$ is a rectangular window of length $M$ in the classical LTE LFDMA, leading to a global filter in the form: \\
\beq
P_k = \left\{
\begin{array}{ll}
	 C_k G_k \hspace{10 mm} &  0 \leq k \leq M-1 \\
	 0 & M \leq k \leq L-1 \\
	\end{array}\right.
\label{equ_overall_p}
\eeq 
The Zero Forcing equalizer for LFDMA writes for $k \in \{0,\hdots, M-1\}$ as follows: \\
\beq 
G_k^{ZF} = \frac{C_k^*}{|C_k|^2}
\eeq  
Thus,   $P_k^{ZF} = 1$ for $ 0\leq k \leq M-1$. \\
As a consequence, the SINR of LFDMA when zero forcing is used becomes: \\
\beq 
SINR_{LFDMA}^{ZF} = \frac{E_s}{N_0} \frac{M^2}{\sum_{k=0}^{M-1}\left|C_k\right|^2 \sum_{k=0}^{M-1}\left|\frac{1}{C_k}\right|^2}
\eeq 
The MMSE equalizer writes as: \\
\beq
G_k^{MMSE} = \frac{C_k^*}{|C_k|^2 +  \frac{1}{M}\left(\frac{E_s}{N_0}\right)^{-1}\sum_{k=0}^{M-1}\left|C_k\right|^2}
\eeq 
The estimated SINR values are derived in the case of a constant channel over one LFDMA symbol, and thus we compare the SINR estimated for block-varying realisations of the pedestrian channel averaged over many SC-FDMA symbols. 
\begin{figure}
\centering
\includegraphics[width=3in]{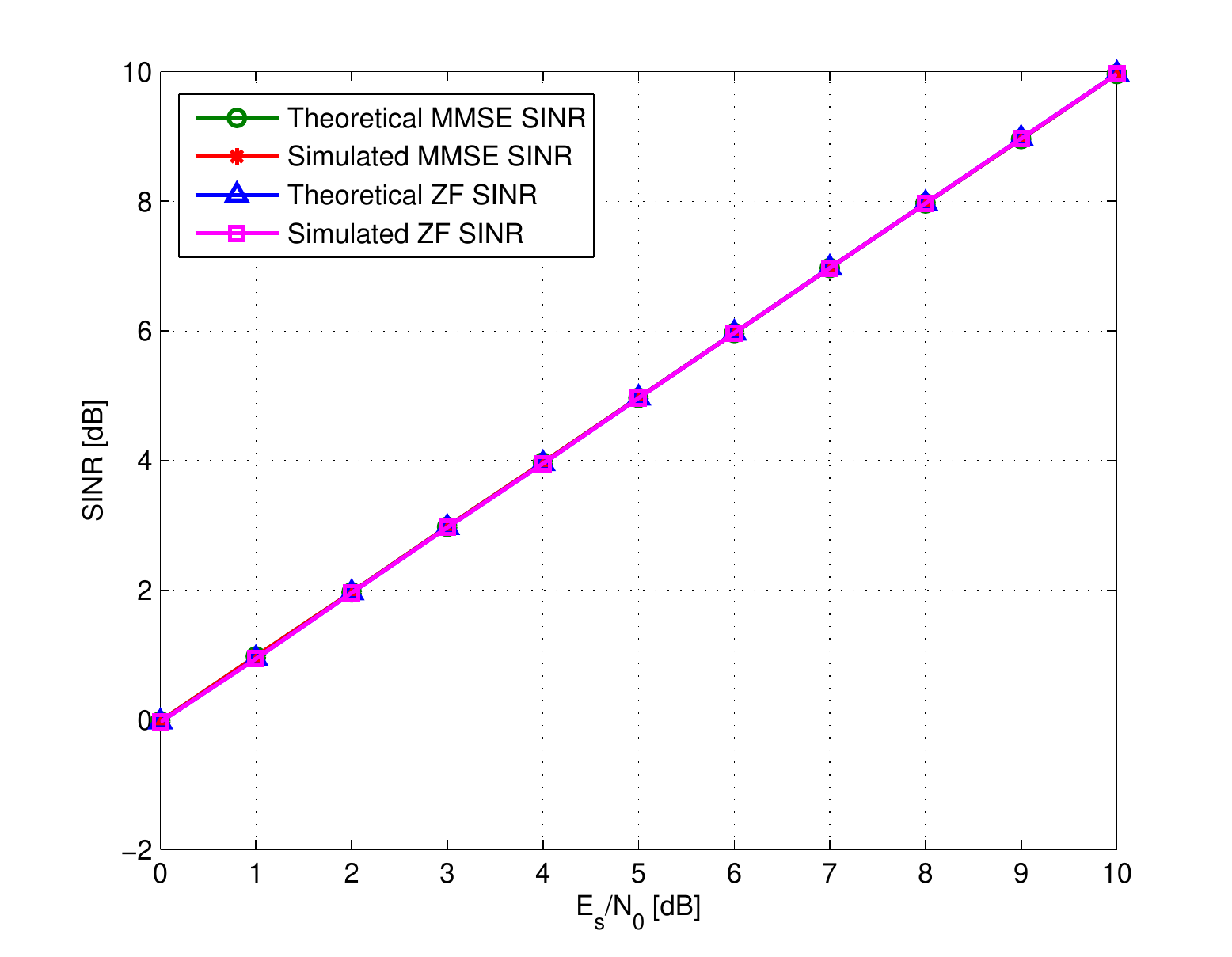}
\caption{MMSE and ZF SINR for fractional rate rectangular LFDMA $N=512$, $M = 10$, and $N_g = 31$}
\label{fig:SINR_ZF_MMSE_rect_frac}
\end{figure}
\begin{figure}
\centering
\includegraphics[width=3in]{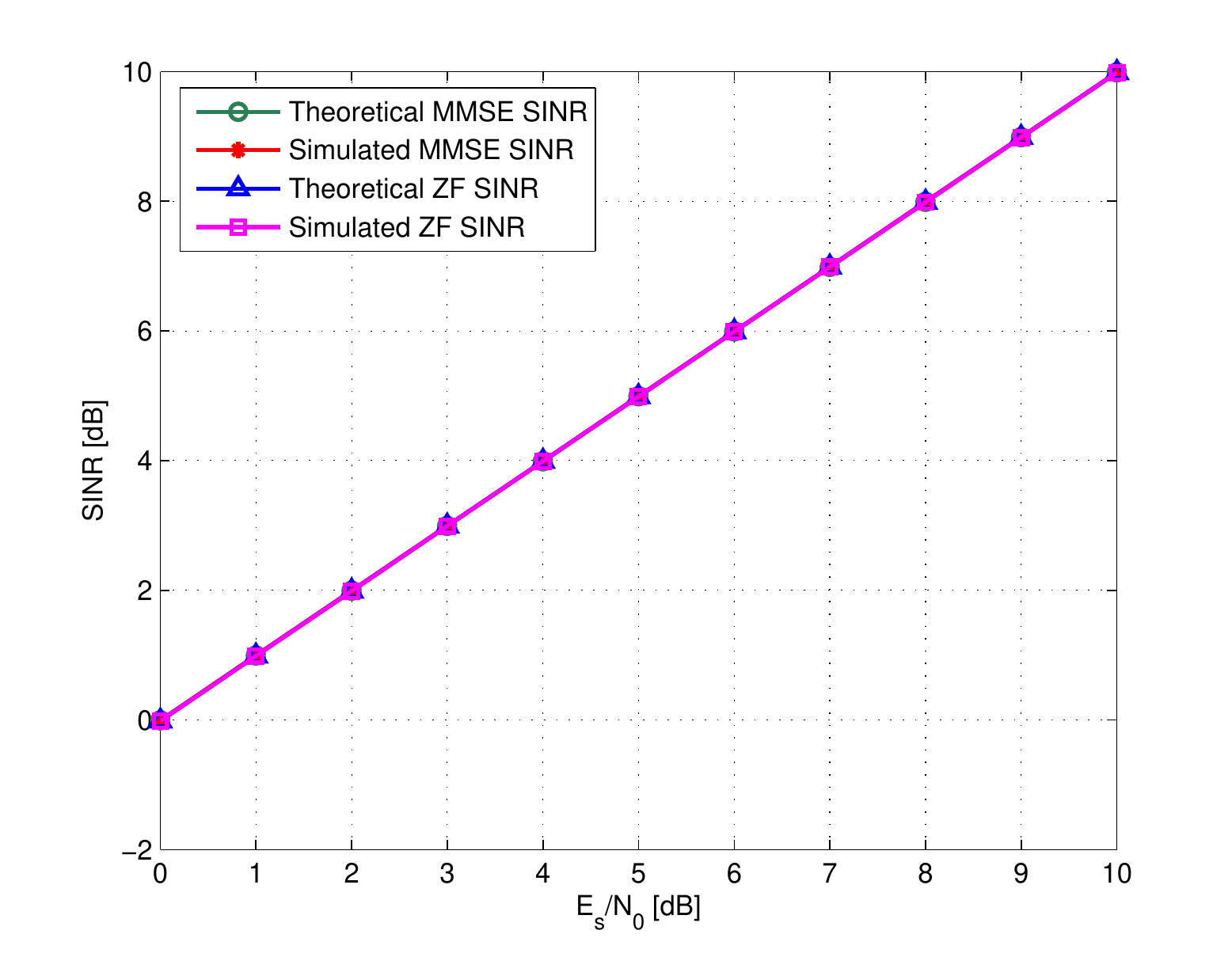}
\caption{MMSE and ZF SINR for fractional rate RRC SS-FDMA $N=512$, $M = 10$, $N_g = 31$, and $\alpha = 0.35$}
\label{fig:SINR_ZF_MMSE_rc_frac}
\end{figure}
 
\rref{fig:SINR_ZF_MMSE_rect_frac} plots the theoretical and estimated SINR for the fractional rate $N=512$ and $M = 10$. 
\subsection{\textbf{SINR of root raised cosine shaped Localised FDMA}}

Let us consider the Root Raised Cosine filter in equ.(\ref{equ:freq_resp_rcf}). It is reminded that the user is then mapped into the second block of frequency bins to avoid border effects with spectral expansion i.e. $\{M-M_{\alpha}, \hdots, 2M+M_{\alpha}-1\}$. In this case, the expression of the overall system frequency response is expressed as follows: 
\beqa 
\sum_{k=0}^{L-1} \left|  G_{k}^{ZF}\right|^2   &= &\sum_{r=0}^{M_{\alpha}-1}  \frac{1}{\left| \vphantom{\sum_{r=0}^{M_{\alpha}-1}} H_{M+r}C_{M+r}\right|^2 + \left|\vphantom{\sum_{r=0}^{M_{\alpha}-1}} H_{2M+r}C_{2M+r} \right|^2} 
 +   \sum_{r=M_{\alpha}}^{M-M_{\alpha}-1}  \frac{1}{\left|\vphantom{\sum_{r=0}^{M_{\alpha}-1}}H_{M+r}C_{M+r}\right|^2} \nnb \\
&+&  \sum_{r=M-M_{\alpha} }^{M-1} \frac{1}{ \left| \vphantom{\sum_{r=0}^{M_{\alpha}-1}}H_{r}C_r\right|^2 + \left| \vphantom{\sum_{r=0}^{M_{\alpha}-1}} H_{M+r}C_{M+r} \right|^2} \nnb
\eeqa 
\rref{fig:SINR_ZF_MMSE_rc_frac} plots the theoretical and estimated SINR for the fractional rate $N=512$ and $M = 10$. 

\section{Conclusion}\label{sec:concl}
In this paper, a  novel time domain implementation of Localised FDMA has been presented. It has been shown that the overall LFDMA system is equivalent to circular convolution with up-sampled time domain filters, in order to address the fractional rate as well. The proposed system allows for general localised mapping schemes including spectral shaping. The PSD of both localised and root raised cosine-spectrally shaped FDMA were investigated and showed the interest of using spectral shaping, as far as side lobes are concerned. SINR formulas for both ZF and MMSE equalizers were derived and confronted to simulation results. The resulting SINR formulas are valuable for system and link level performance evaluation using Frame Error Rate prediction, which were out of the scope of this paper.
\section{Acknowledgements}
Part of this study was carried out in a R \& T action (R-S12/TC-0008-002) at the French Space Agency (CNES). 
\section{APPENDIX}

\subsection{Equivalent noise variance} 
The equivalent noise issues from up-sampling by a factor $L_N$ equalizing then down-sampling by a factor $L_M$. This results in a cyclo-stationary equivalent noise, the autocorrelation of which reads as:  \\
\beq 
 R_{\tilde{w}}(k,n) = E[\tilde{w}_k \tilde{w}_{k-n}^*] \nnb = L_M^2 \sigma_w^2 \sum_{m=0}^{N-1} g(<kL_M - mL_N>_L) g^*(<(k-n)L_M -mL_N>_L) \nnb  \\ 
\eeq
We can show that this autocorrelation function is $L_N$-periodic in time by expressing it in the frequency domain as follows:  \\ 
\beqa
 R_{\tilde{w}}(k,n) &=& \frac{\sigma_w^2}{M^2}  \sum_{m=0}^{N-1} \sum_{i=0}^{L-1}  \sum_{i'=0}^{L-1} G_i G_{i'}^* \Omega_{L}^{i(kL_M - mL_N)}\Omega_{L}^{-i'((k-n)L_M -mL_N)} \nnb \\
&=& \frac{\sigma_w^2}{M^2}  \sum_{i=0}^{L-1} \sum_{i'=0}^{L-1} G_i G_{i'}^* \Omega_{L}^{k(i-i')L_M} \Omega_{L}^{i'n L_M } \sum_{m=0}^{N-1}\Omega_{N}^{m(i-i')} \nnb \\
 &= & \frac{\sigma_w^2 N}{M^2}  \sum_{r=0}^{N-1} \sum_{s=0}^{L_N-1} \sum_{s'=0}^{L_N-1} G_{sM+r}G_{s'M+r}^*\Omega_{L}^{k(s-s')L_M N}  
 \Omega_{L}^{(s'N+r)nL_M} \nnb \\
&=& \frac{\sigma_w^2 N}{M^2}  \sum_{r=0}^{N-1} \sum_{s=0}^{L_N-1} \sum_{s'=0}^{L_N-1} G_{sM+r}G_{s'M+r}^* \Omega_{L}^{k(s-s')L_M N} \Omega_{L}^{(s'N+r)nL_M} \nnb
\eeqa
We used the fact that $L_N$ is the least integer $A$ such that $A N = B M$ since it is obtained from the least common multiple of $M$ and $N$. \\ 
Thus we compute the stationary power spectral density of the noise, by averaging over the period $L_N$ as follows: \\
\beqa
\overline{R}_{\tilde{w}}(n)= \frac{1}{L_N} \sum_{k=0}^{L_N-1}R_{\tilde{w}}(k,n) &= &
  \frac{\sigma_w^2 N}{M^2 L_N}  \sum_{r=0}^{N-1} \sum_{s=0}^{L_N-1} \sum_{s'=0}^{L_N-1} G_{sM+r}G_{s'M+r}^* \sum_{k=0}^{L_N-1} \Omega_{L_N}^{k(s-s')L_M} \Omega_{L}^{(s'N+r)nL_M} \nnb \\
&=& \frac{\sigma_w^2 N}{M^2} \sum_{r=0}^{N-1} \sum_{s=0}^{L_N-1} G_{sM+r}G_{sM+r}^*\Omega_{L}^{(sN+r)nL_M} \nnb \\
& = &\frac{\sigma_w^2 N}{M^2} \sum_{r=0}^{N-1}\left|G_k \right|^2\Omega_{M}^{kn}  
\eeqa 
where the transition from the $1^{st}$ to the $2^{nd}$ equality results from the fact that $L_M$ and $L_N$ are coprime. As a result, the stationarized noise variance is: \\
\beq 
\sigma_{\tilde{w}}^2 =  \frac{\sigma_w^2 N}{M^2} \sum_{r=0}^{N-1}\left|G_k \right|^2 
\label{equ:noise_variance_appendix}
\eeq
It can be noticed that stationarizing the noise would not be necessary if the equalization function $G$ had only $N$ non zero values i.e. $G_{r} = 0 \: if \: r \geq N $. This would result in a stationary noise which covariance is similar to (\ref{equ:noise_variance_appendix}). 
\bibliographystyle{IEEEbib}
\bibliography{sc_fdma}

\end{document}